\documentclass{WileyMSP-template}
\usepackage{graphicx}
\usepackage{soul}
\usepackage{amsmath}
\usepackage{amsfonts}
\usepackage{amssymb}
\usepackage{color}
\usepackage{float}
\usepackage{verbatim}
\usepackage{stackrel}
\usepackage[colorlinks,linkcolor=blue,citecolor=blue,urlcolor=blue]{hyperref}
\usepackage[capitalize]{cleveref}
\usepackage{tikz}
\usepackage{bm}
\usepackage{ragged2e}
\DeclareMathAlphabet\mathbfcal{OMS}{cmsy}{b}{n}
\justifying
\graphicspath{{./Figs/}}
\begin{document}

\pagestyle{fancy}
\rhead{\includegraphics[width=2.5cm]{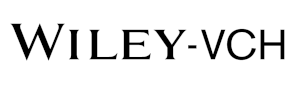}}

\title{Time resolved optical response of the Dicke's model via the nonequilibrium
Green's function approach}

\maketitle

\author{Megha Gopalakrishna*,}
\author{Yaroslav Pavlyukh*,}
\author{Claudio Verdozzi*}

\begin{affiliations}
\noindent Megha Gopalakrishna\\
Department of Physics, Division of Mathematical Physics,\\
Lund University, 22100 Lund, Sweden\\
Email Address: megha.gopalakrishna@teorfys.lu.se\\

\noindent Yaroslav Pavlyukh\\
Institute of Theoretical Physics, Faculty of Fundamental Problems of Technology,\\
Wroclaw University of Science and Technology, 50-370 Wroclaw, Poland\\
Email Address: yaroslav.pavlyukh@pwr.edu.pl\\ 

\noindent Claudio Verdozzi\\
Department of Physics, Division of Mathematical Physics and ETSF,\\
Lund University, 22100 Lund, Sweden\\
Email Address: Claudio.Verdozzi@teorfys.lu.se\\

\end{affiliations}


\begin{abstract}
Due to their conceptual appeal and computational convenience, two-level systems (TLS) and their generalisations are
often used to investigate nonlinear behavior in quantum optics, and to assess the applicability of theoretical
methods. Here the focus is on second harmonic generation (SHG) and, as system of interest, on the Dicke model, which
consists of several TLSs inside an optical cavity. The main aspect addressed is the scope of non-equilibrium Green's
function (NEGF) to describe the effect of disorder and electron-electron (e-e) interactions on the SHG signal.  For
benchmarking purposes, exact diagonalization (ED) results are also presented and discussed.  SHG spectra obtained with
NEGF and ED are found to be in very good mutual agreement in most situations. Furthermore, inhomogeneity in the TLS and
e-e interactions reduce the strength of SHG, and the reduction is stronger with inhomogeneity than with
interactions. This trend is consistently noted across different (small to large) system sizes. Finally, a modified NEGF
approach is proposed to account for cavity leakage, where the quantum photon fields are coupled to a bath of classical
oscillators. As to be expected, within this mixed quantum-classical scheme a decrease in the intensity of the
fluorescent spectra takes place depending on the entity of cavity leakage.
\end{abstract}

 \section{Introduction}
 
Second harmonic generation (SHG) is a non-linear phenomenon in which two incident photons are absorbed simultaneously,
and a single photon is emitted with doubled frequency \cite{BloeRMP}. SHG is the focus of extensive basic and applied
research in physics, chemistry, biology, engineering and medicine \cite{physi,engine,chemist,biolog,medicine}, because
of its fundamental interest as optical process, and because its vast use in characterization technique and optical
devices. Extensive investigations of SHG have been conducted both theoretically and experimentally in many systems and
there is now a rather comprehensive and firm understanding of this optical process. There are, however, still some
particular situations where our grasp of SHG remains conceptually not complete. The one of interest in this work is when
many photon effects play a role, and yet quantum fluctuations remain noticeable \cite{Cini1993,Cini1995,Emil2020}. This
paper focusses on this specific regime, using the Dicke model \cite{Dicke1954} as the system of interest.

The Dicke model holds significance in cavity optics, and has been vastly used in the study of superradiance
\cite{Dicke1954,Kirton2019}. It is also a go-to model for many theoretical explorations concerning basic aspects of
light-matter interaction, and has also been considered in a simplified form by applying the so-called rotating wave
approximation (RWA)\cite{RWA1,RWA2}: in this case, the model is mostly known as as the Tavis-Cummings model
\cite{TavisCummings} (see e.g. \cite{Keeling2009} for a review).

In more detail, the original Dicke model \cite{Dicke1954,Kirton2019} describes a cavity mode interacting
  with a series of identical two-level systems (TLSs), and the original Dicke Hamiltonian is
\begin{align}\label{originalDicke}
  \hat H_{DM}=\omega_a \hat a^\dagger \hat a
  + \Delta\sum_{{ \bm i}=1}^L{\hat s^z_{ \bm i}}
  + 2g_a (\hat a^\dagger +\hat a)\sum_{\bm{i}=1}^L\hat s^x_{ \bm i},
\end{align}
where $L$ represents the number of TLSs in the system, $\Delta$ is the difference in energy between ground and excited
level in a two level systems (TLS), and $\hat a^\dagger$ creates a photon in the cavity field. Furthermore, the cavity
mode frequency is { $\omega_a$}, $g_a$ represents the coupling between a TLS and the cavity field (in the original paper
\cite{Dicke1954}, $g_a\equiv \lambda/\sqrt{L}$, where $\lambda$ is used to characterise the strength of the
electron-photon coupling), {\color{black} $\hat s^{j}_{\bm i}=\frac12
  \sum_{\tau\tau'}\hat{c}_{\tau,\bm{i}}^\dagger\sigma^{j}_{\tau\tau'}\hat{c}_{\tau',\bm{i}}$ are the spin operators,
  $\sigma^{j}$ are the Pauli spin matrices ($j=x,y,z$), and $\hat{c}_{\tau,\bm{i}}^\dagger$, $\hat{c}_{\tau,\bm{i}}$ are
  the electron creation and annihilation operators in $\bm{i}$th TLS. Index $\tau$ takes values $g$ (ground state) or
  $e$ (excited state).}

Since we aim to investigate the SHG response from the Dicke model, the Hamiltonian in Eq.~(\ref{originalDicke}) will be
augmented with a second photon mode representing a fluorescent field. Several studies devoted to the Dicke model address
the homogeneous case, where all the TLS are identical. There has however been also considerable interest in studying
disordered Dicke systems, as they are relevant in describing situations such as cold atoms in a cavity
\cite{Eastham2001,Tsyplyatyev2009}. In this respect, another way to enrich the physics of the model is to consider the
interaction between th electrons to make the system more interesting.

The inclusion of interactions in the Dicke model has received some attention in the literature, for example in the
context of quantum information via the so-called Hubbard-Dicke model for interacting qubits in coupled cavities (see
e.g. \cite{ShuHeQPT2022}), or interacting Rydberg atoms is optical cavities (see e.g. \cite{PRLRydberg2013}). In
situations of this kind, mostly spin-spin type (for dipole-dipole, qubit-qubit) interactions are considered between the
relevant (atom-atom, qubit-qubit, etc) units.  This choice is not necessarily always physically motivated, but in any
case it makes these extended models amenable to simplified numerical treatments.

The effect of both disorder and interactions on SHG in the Dicke model is the target of the present
  investigation and we will consider a more general type of interactions, namely Hubbard-type interactions between
  electrons in nearest neighbour TLSs.

As method of choice to tackle large samples with disorder and this type of interactions, we advocate the use of the
non-equilibrium Green's functions (NEGF) method \cite{Kadanoff,Keldysh,Balzer2012,GSRvLbook,Sakkinen2015} within the
so-called generalized Kadanoff-Baym ansatz (GKBA) \cite{Lipavsky86}, and in a formulation apt to deal with
electron-boson systems \cite{Karlsson2021,Pavlyukh_Mar_2022,Pavlyukh_Nov_2022}.  Furthermore, to have a most transparent
platform to assess the scope of NEGF-GKBA, we will consider for simplicity only interactions between electrons in the
excited levels of the TLS.  We will see in the following that this model already represents a valuable benchmark for
NEGF-GKBA.

After introducing the model in detail in Sect.~\ref{Xmodel}, we briefly review the NEGF-GKBA method in
Sect.~\ref{negf_method}, and then we proceed to a comparison of NEGF-GKBA results with some exact diagonalization {ED}
benchmarks in Sect.~\ref{comparazioni}.  This is followed by a discussion of interaction and disorder effects on SHG in
Sect.~\ref{disordinato}, and, in in Sect.~\ref{perdite}, by an analysis of how cavity leakage from an imperfect cavity
manifests in the SHG spectra.  Finally, after a brief discussion of the third-harmonic generation response in
Sect.~\ref{terza}, we conclude in Sect.~\ref{finito} with some conclusive remarks and a succinct outlook.

\section{Model Hamiltonian}\label{Xmodel}
For convenience, we
rewrite $\hat H_{DM}$ of Eq.~(\ref{originalDicke})
in an explicit electron picture (in the rewriting, an obvious constant energy shift of the Hamiltonian is neglected):
 \begin{align}
   \hat H_{DM}\rightarrow \hat{\tilde{H}}_{DM}=\epsilon_g \sum_{\bm{i}=1}^L \hat c_{g,\bm i}^{\dagger} \hat c_{g,\bm i}
   +\epsilon_e \sum_{\bm{i}=1}^L \hat c_{e,\bm i}^{\dagger}\hat c_{e,\bm i}+\omega_a \hat a^\dagger \hat a
   +g_a(\hat a^\dagger+\hat a)\sum_{\bm {i}=1}^L(\hat c_{e,\bm i}^{\dagger}\hat c_{g,\bm i}+\hat c_{g,\bm i}^{\dagger} \hat c_{e,\bm i}),
 \end{align}
where $\epsilon_g\,(\epsilon_e)$ corresponds to the ground (excited) level energy in a TLS, $\hat c_{g/e,\bm i}$
destroys an electron in the ground/excited level in $\bm{i}$th TLS (at all times, there is only one electron in each
TLS, and $\epsilon_e-\epsilon_g=\Delta$).
     
In this work we intend to explore the second harmonic generation (SHG) in a Dicke system in the presence of e-e
interactions and energies of the TLSs are inhomogeneously distributed. To this end, $\hat{\tilde{H}}_{DM}$ is augmented
with a second photon mode, accounting for the fluorescent field. Furthermore, to take into account e-e interactions, we
consider the TLSs as arranged and numbered according to a given sequence, and we use a notion of ``adjacency'' between
TLS with consecutive numbers. With these extensions and conventions, the system's Hamiltonian $\hat{\tilde{H}}_{DM}
\rightarrow \hat{H}_S $, with
 \begin{equation}
  \hat{H}_S=\hat{H}_{ele}+\hat{H}_{ph}+\hat{H}_{e-ph}.
  \label{sys_Ham}
 \end{equation}
We now specify in detail the different terms in $\hat{H}_S$, starting from the electronic part. This reads
\begin{align}
  \hat{H}_{ele}= \sum_{\bm{i}}^L \tilde{\epsilon}_{g, \bm i} \hat c_{g, \bm i}^\dagger \hat c_{g, \bm i}
  +\sum_{\bm{i}}^L\tilde{\epsilon}_{e, \bm i} \hat{c}_{e, \bm i}^\dagger \hat{c}_{e, \bm i}
  + \sum_{<\bm i,\bm j>}^L \frac{U_e}{2} \hat{n}_{e, \bm i} \hat{n}_{e, \bm j},
 \label{H_ele}
\end{align}
The electronic interactions are introduced between excited electrons of TLSs with nearest neighbour sites indexes ${ \bm
  i, \bm j}$ (denoted by the $\langle {\bm i, \bm j}\rangle$), with $\hat{n}_{e,\bm i}=\hat c_{e, \bm i}^{\dagger}\hat
c_{e, \bm i}$ the density operator of the excited state of the ${ \bm i}$-th TLS, and $U_e$ the strength of the
  interaction.  A similar contribution could be considered for the ground levels as well, but is omitted here for
  simplicity, and to focus on the role of interactions in the system's excited states.  The effect of inhomogeneity in
  the TLSs is included by choosing the ground and the excited level energies as
\begin{align}\label{periodis}
 &\tilde{\epsilon}_{{ g, \bm i}}=\epsilon_g-\frac{\delta}{2}\sin\Big[\frac{\pi({ \bm i}-1)}{L}\Big]\\
 &\tilde{\epsilon}_{{ e, \bm i}}=\epsilon_e+\frac{\delta}{2}\sin\Big[\frac{3\pi({ \bm i}-1)}{L}\Big].
\end{align}
These energy distributions will introduce a pseudo-disorder (in the following, just referred to as ``disorder'' for
simplicity) in the resonant frequencies of the Dicke model, meaning that many TLSs are actually off-resonance whenever
$\delta \neq 0$, at an extent determined by the value of $\delta$.

The free term for the two photon fields are
\begin{equation}
 \hat H_{ph}=\omega_a \hat a^\dagger \hat a+ \omega \hat b^\dagger \hat b,
\end{equation}
where $\hat b$ destroys a fluorescent photon, whilst  the interaction between the electrons and the photons is represented by
\begin{equation}
  \hat{H}_{e-ph}=\left[g_a(\hat a^\dagger+\hat a)+g_b(t)(\hat b^\dagger+\hat b)\right]
  \sum_{{ \bm i}}^L(\hat c_{{g, \bm i}}^{\dagger}\hat c_{{ e, \bm i}} +\hat c_{ e, \bm i}^{\dagger}\hat c_{ g, \bm i}),
\end{equation}
where $g_b(t)=g' e^{-\Gamma t}$. Here, $g'$ is the initial strength of the coupling between the electron and the
fluorescent field, and $\Gamma$ accounts via a phenomenological damping (broadening) effects left out in the model,
e.g., non-radiative transitions and/or mode leakages in the cavity \cite{Cini1993,Emil2020,Scipost}. In what follows, we
introduce the generalised indices $(g,\bm{j}) \rightarrow 2j-1$ and $(e,\bm{j})\rightarrow 2j$.

\section{The NEGF method}\label{negf_method}
A straightforward method to describe the behavior of our modified Dicke Hamiltonian is exact diagonalization (ED), which
is however limited to rather small samples.  Nevertheless, if one adopts a spin formalism for the Dicke model (i.e. each
TLS is represented as a spin 1/2 degree of freedom, then in the absence of interaction and disorder the Hamiltonian
commutes with the total spin of the system. In that case, starting from a state of definite total spin, the exact
numerical time evolution can be performed by considering only a spin-conserved subspace of the complete basis set. With
this reduction, the capacity of ED can be stretched to study larger Dicke systems. On the other hand, with interaction
and disorder present (in the sense specified in Eq.~\ref{sys_Ham}) the total spin is not conserved, and ED can still be
employed in the electronic-level picture, but not for large samples.

This brings us to the central point of the paper, namely the proposal that for large samples with disorder and/or
interactions, we can make use of the NEGF-GKBA in a formulation suitable for electron-boson systems
\cite{Karlsson2021,Pavlyukh_Mar_2022,Pavlyukh_Nov_2022}.

Recently, a formulation of the NEGF-GKBA method was introduced \cite{Schlunzen2020,Joost2020}, which attains time linear
scaling in the simulation time. The approach has subsequently been extended to case of interacting electron-boson
systems \cite{Karlsson2021,Pavlyukh_Mar_2022,Pavlyukh_Nov_2022}. In this work, we consider this linear-time formulation
to study Dicke systems of large size and in the rest of this section we provide a short summary of the method, referring
to the original literature for details.

In the theoretical treatment, the photon fields are represented by position and momentum operators instead of
creation/annihilation operators. For the cavity field, the position and momentum operators are $\hat x_1=\nobreak(\hat
a^\dagger+ \nobreak \hat a)/\sqrt{2}$ and $\hat p_1=i(\hat a^\dagger- \nobreak \hat a)/\sqrt{2}$. Similarly, the
fluorescent field is represented by $\hat x_2$ and $\hat p_2$. Furthermore, it is useful to represent jointly the
position and momentum operators via an operator $\hat \phi_\mu$, where the index $\mu$ is a combined index with $\hat
\phi_{i,1}=\hat x_i$ and $\hat \phi_{i,2}=\hat p_i$. Hence, in terms of the $\{\hat \phi_\mu\}$ operators, the
Hamiltonian of the photon fields results in $\hat H_{ph}=\sum_{\mu\nu}\Omega_{\mu\nu}\hat \phi_{\mu}\hat \phi_\nu$ and,
for the electron-photon interaction Hamiltonian, we can write, in the same notation, $\hat
H_{el-ph}=\sum_{\mu,ij}g_{\mu,ij}(t)\hat c_i^\dagger \hat c_j\hat \phi_{\mu}$.

In the NEGF-GKBA scheme for electron-boson systems, the central ingredient is the pair of density matrices $(\rho,
{\boldsymbol\gamma})$, respectively for electrons and for bosons
\cite{Karlsson2021,Pavlyukh_Mar_2022,Pavlyukh_Nov_2022}. More in detail, the (single-particle) density matrix for
electrons is $\rho_{ij}(t)=\big<\hat c_j^\dagger(t) \hat c_i(t) \big>$, while the boson density matrix is given by
$\gamma_{\mu\nu}(t)=\big<\Delta\hat \phi_\nu(t)\Delta\hat \phi_\mu(t)\big>$, where $\Delta\hat \phi_\nu(t)\equiv\hat
\phi_\nu(t)-\phi_\nu(t)$ and {\color{black} $\phi_\nu(t)\equiv\langle\hat\phi_\nu(t)\rangle$}. In terms of $\rho$ and
${\boldsymbol\gamma}$ it is then convenient to consider an effective one-particle electronic Hamiltonian $h^e$ at the
Hartree-Fock {(HF)} level,
\begin{align}
h^e_{ij}(t)=h^0_{ij}+\sum_m\delta_{ij}\mathcal{U}_{im}\rho_{mm}^<(t)-\mathcal{U}_{ij}\rho_{ij}^<(t) +\sum_\mu\phi_\mu(t)g_{\mu,ij}(t),
\end{align}
where $h^0_{ij} \hat c_i^\dagger \hat c_j$ denotes all the non interacting contributions in Eq.~(\ref{H_ele}), and the
interaction terms are chosen as $\mathcal{U}_{ij}=U_{ij}\delta_{i,j\pm1}$ (the appropriate choice of $U_{ij}$ permits to
specialise the type of interactions, e.g. only between excited levels).  The GKBA for fermions has been scrutinised
quite in detail in the literature. On the other hand, in this work we are primarily interested on the scope of NEGF-GKBA
for SHG in the multi-photon regime and for models of quantum optics. This, together with our wish of avoiding
unnecessarily heavy calculations, motivates our simple, HF treatment of e-e interactions.  In the same spirit, the
effective boson Hamiltonian $h^b_{\mu\nu}=\nobreak2\sum_{\xi}\alpha_{\mu\xi}\Omega_{\xi\nu}$ with
{$\alpha_{\mu\nu}=[\hat \phi_\mu,\hat \phi_\nu]$}.

The resulting equations of motion for the density matrices assume then the form
\cite{Pavlyukh_Mar_2022,Pavlyukh_Nov_2022}
\begin{align}
 \frac{d}{dt}\rho^<(t)&=-i\big[h^e(t),\rho^<(t)\big]-(I^e(t)+{I^e}^\dagger(t)),\\
 \frac{d}{dt}{\boldsymbol\gamma}^<(t)&=-i\big[{\boldsymbol h}^b(t),{\boldsymbol \gamma}^<(t)\big]
 +({\boldsymbol I}^b(t)+{{\boldsymbol I}^{b}}^\dagger(t))
\end{align}
where the collision integrals $I^{e}$ and ${\boldsymbol I}^b$ account for the interaction between electrons and photon
fields.  More explicitly, $I^e_{mj}=i\sum_{\mu,l}g_{\mu,ml}\mathcal{G}^b_{\mu,lj}$ and
$I^b_{\mu\nu}=-i\sum_{\eta,mj}\alpha_{\mu\eta}g_{\eta,mj}\mathcal{G}^b_{\nu,jm}$, where the high-order Green's function
$\mathcal{G}^b_{\mu,ij}(t)=\big<\hat c^\dagger_j(t)\hat c_i(t)\hat \phi_\mu(t)\big>_c$.

To explicitly compute the collision integrals within many-body perturbation theory, use is then made
of the electron and the boson GKBA scheme: 
\begin{align}
 G^{\lessgtr}(t,t')&=-G^R(t,t')\rho^{\lessgtr}(t')+\rho^{\lessgtr}(t)G^A(t,t')\\
 {\boldsymbol{D}}^{\lessgtr}(t,t')&={\boldsymbol D}^R(t,t'){\boldsymbol\alpha \boldsymbol \gamma}^{\lessgtr}(t')
 -{\boldsymbol \gamma}^{\lessgtr}(t){\boldsymbol \alpha} {\boldsymbol D}^A(t,t'),
\end{align}
where $G$ denotes the single particle electron Green's function, ${\boldsymbol D}$ is the corresponding boson
counterpart, $R$ ($A$) denote the retarded (advanced) components, and the connection to the density matrices is given by
$G^\lessgtr(t,t)=i\rho^\lessgtr(t)$ and ${\boldsymbol D}^\lessgtr(t,t)=-i{\boldsymbol \gamma}^\lessgtr(t)$.  In terms of
the GKBA expression above, a time linear scheme is obtained, where the ensuing coupled ordinary differential equations
read:
\begin{subequations}
\begin{align}
i\frac{d}{dt}\phi_\mu&=\sum_\nu h^b_{\mu\nu}\phi_\nu + \sum_{\nu,ij}\alpha_{\mu\nu}g_{\nu,ij}\rho^<_{ji} \\
 i\frac{d}{dt}\rho_{lj}^<&=[h^e,\rho^<]_{lj}+\sum_{\mu,i}g_{\mu,li}\mathcal{G}_{\mu,ij}^b-(l\leftrightarrow j)^*,\\ 
 i\frac{d}{dt}\gamma_{\mu\nu}^<&=[{\boldsymbol h}^b,{ \boldsymbol \gamma}^<]_{\mu\nu}
 +\sum_{\eta,mn}\alpha_{\mu,\eta}g_{\eta,mn}\mathcal{G}^b_{\nu,nm}-(\mu\leftrightarrow \nu)^*,\\
 i\frac{d}{dt}{ \mathbfcal{G}}^b&=-{\boldsymbol \Psi}^b+{ \boldsymbol h}^b{\mathbfcal{G}}^b-{\mathbfcal{G}}^b{\boldsymbol h}^e,
\end{align}
\label{NEGF-GKBA}
\end{subequations}
where ${\boldsymbol h^e=h^e\otimes I-I \otimes (h^e)^T}$, ${\boldsymbol\Psi}^b\equiv{\boldsymbol \gamma}^>{\boldsymbol
  g}{\boldsymbol \rho}^<-{ \boldsymbol \gamma}^< { \boldsymbol g} { \boldsymbol \rho}^>$ with ${ \boldsymbol
  \rho}^>=\rho^>\otimes(\rho^<)^T$ and ${ \boldsymbol \rho}^<=\rho^<\otimes(\rho^>)^T$.

Within this NEGF-GKBA treatment, we use the average number of emitted photons $\big<\hat b^\dagger \hat b\big>$ as
indicator of the occurrence of SHG and, as the system's initial state, we take a product state of the electron system
and the two photon fields. As in { e.g.\cite{Cini1993,Emil2020,PhysRevB.105.125135,Scipost}}, initially the electron
system will be in the ground state, whilst the initial state of the cavity field is a coherent state
\begin{align}
\color{black}\left|\beta\right\rangle & \color{black}= e^{\beta \hat a^\dagger- \beta^*  \hat a}|0\rangle,
\end{align}
and the emitted photon field is initially represented by the vacuum state $\left|0\right\rangle$. In all
the calculations, the resonance frequency of a TLS without disorder is taken ${\Delta}=\epsilon_e-\epsilon_g=1.0$, the
cavity field coupling is $g_a=0.03$, and the fluorescent field coupling $g'=0.01$, with $\Gamma=0.02$.  The coherent
state of the cavity field is chosen to contain { $n_a=\langle \hat a^\dagger \hat a\rangle=\beta^2=9$} average number of
photons.  Finally, the cavity frequency in all the SHG calculations is ${\omega_a}={\Delta}/2=0.5$.


\section{ED and NEGF-GKBA}\label{comparazioni}
\begin{figure*}
 \begin{tikzpicture}
  \node at ( 0.0, 2.0) {\includegraphics[width=0.34\columnwidth]{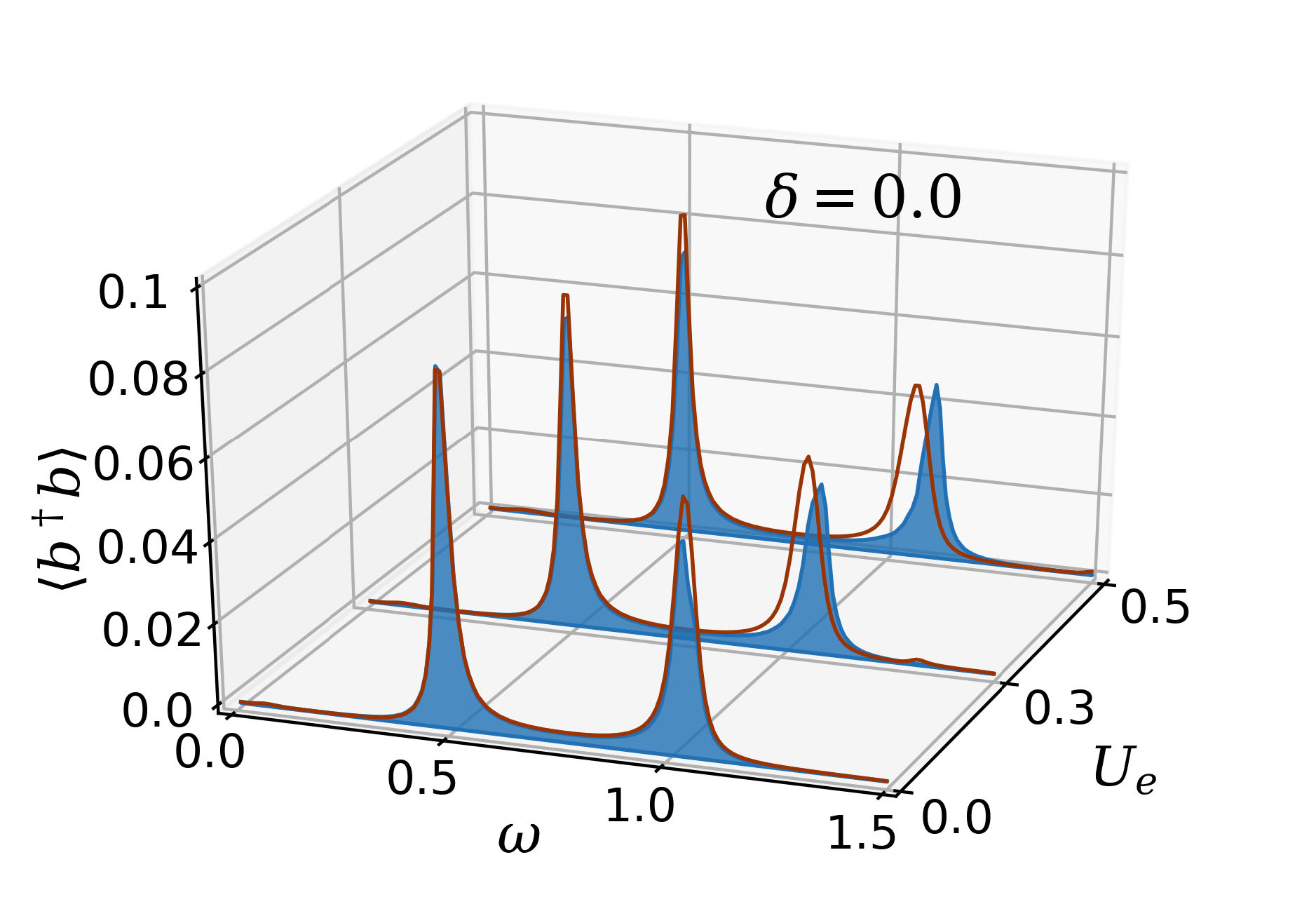}}; 
  \node at ( 6.0, 2.0) {\includegraphics[width=0.34\columnwidth]{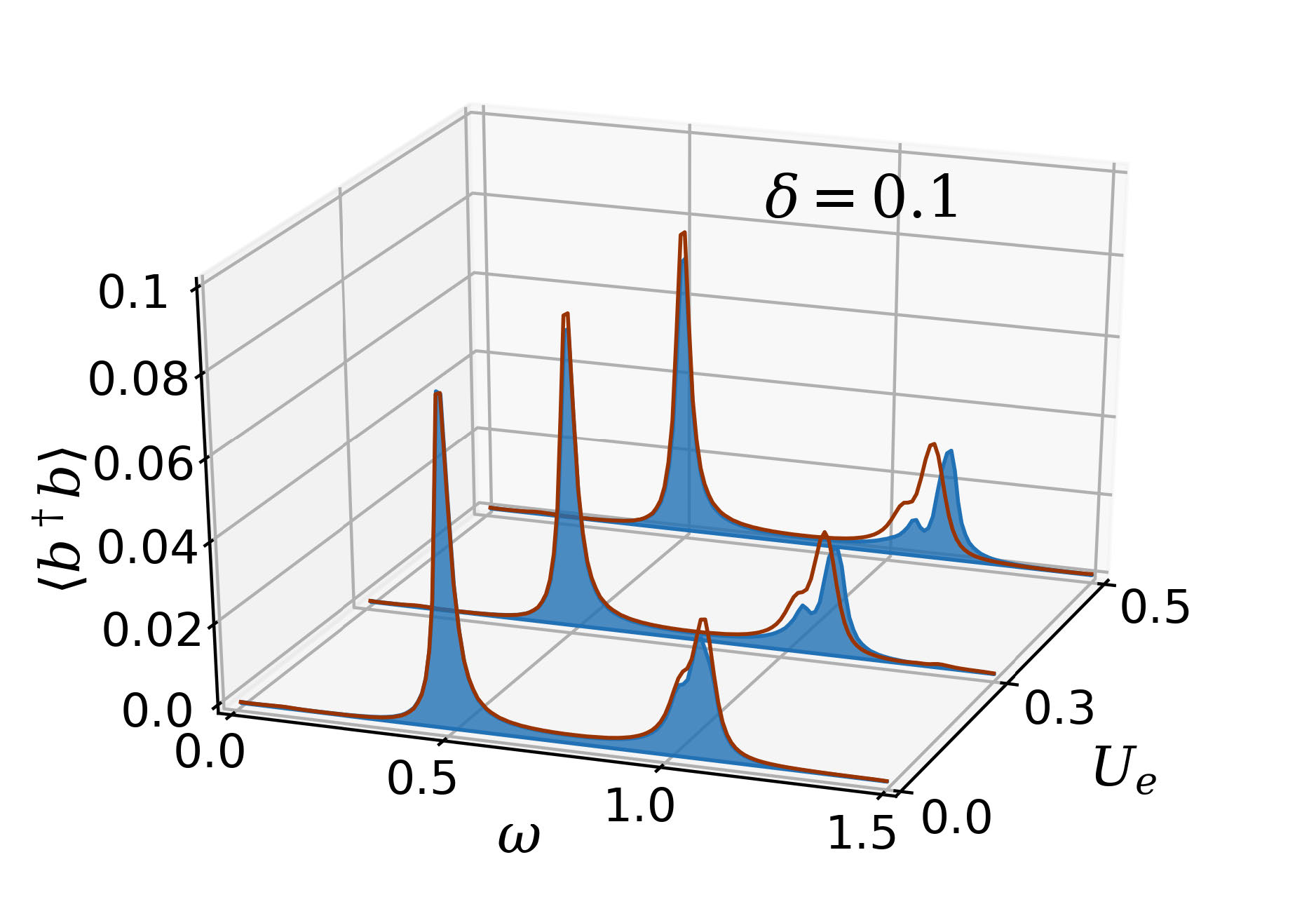}};
  \node at (12.0, 2.0) {\includegraphics[width=0.34\columnwidth]{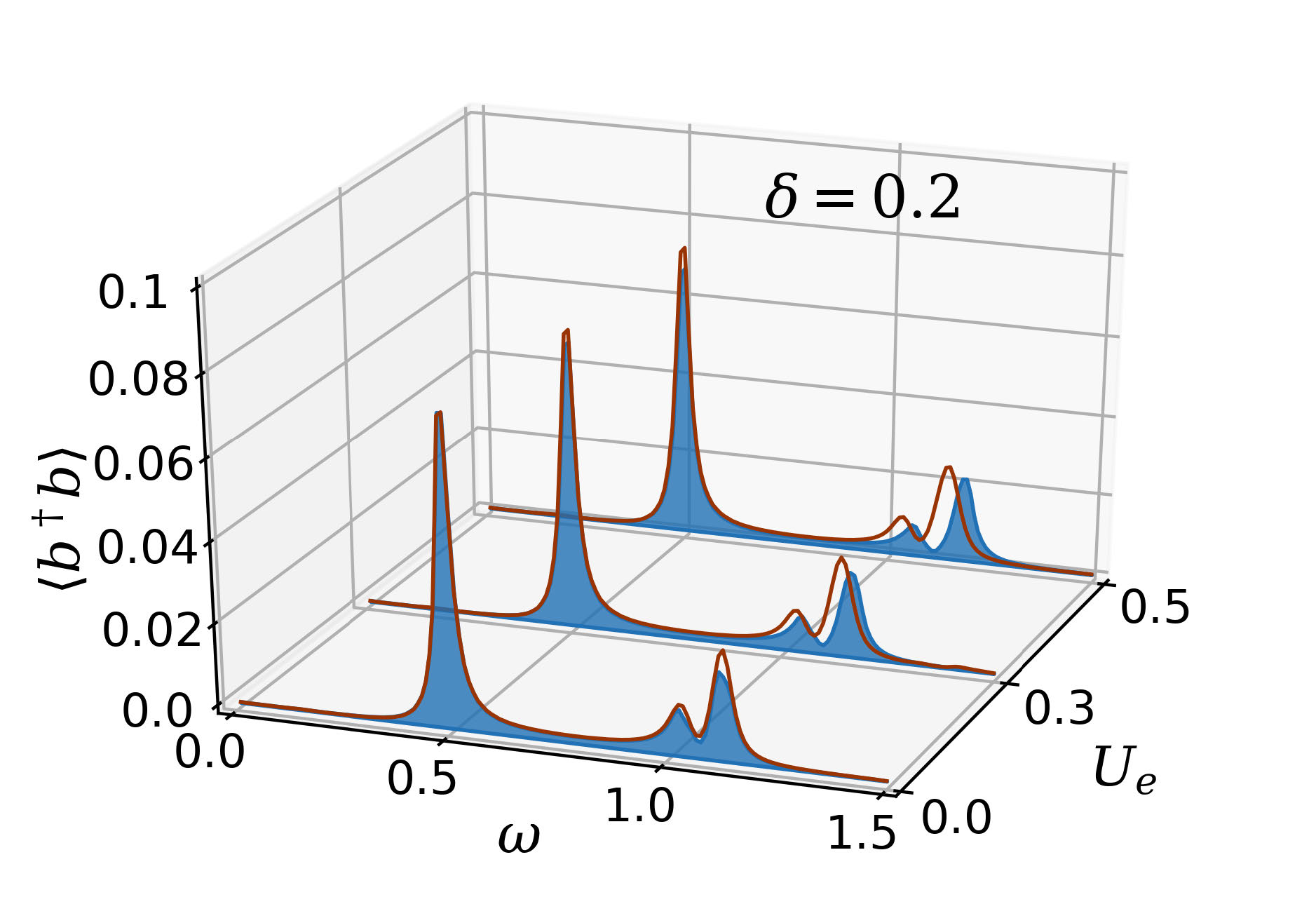}};
  \node at ( 0.0,-2.50) {\includegraphics[width=0.34\columnwidth]{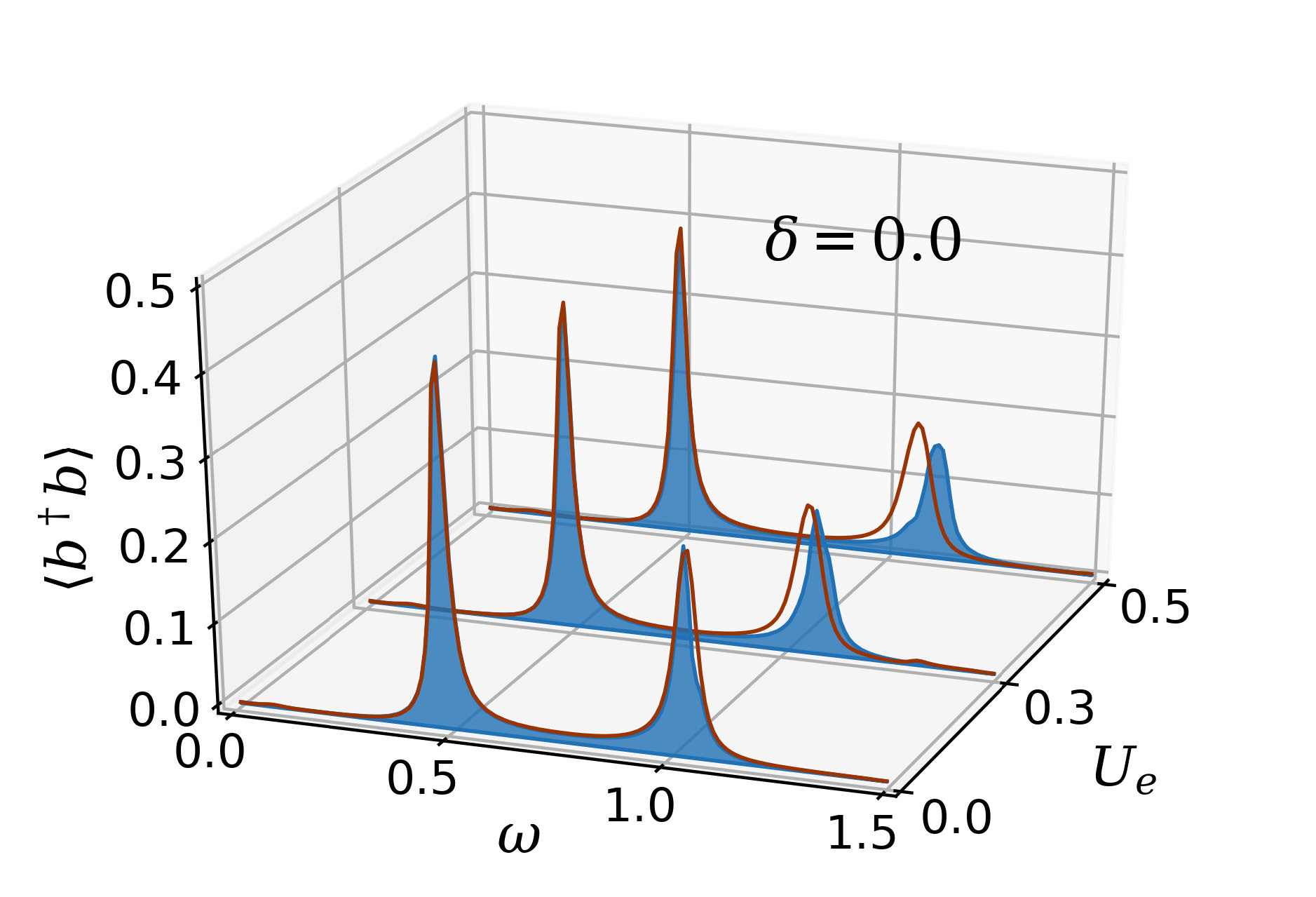}};
  \node at ( 6.0,-2.50) {\includegraphics[width=0.34\columnwidth]{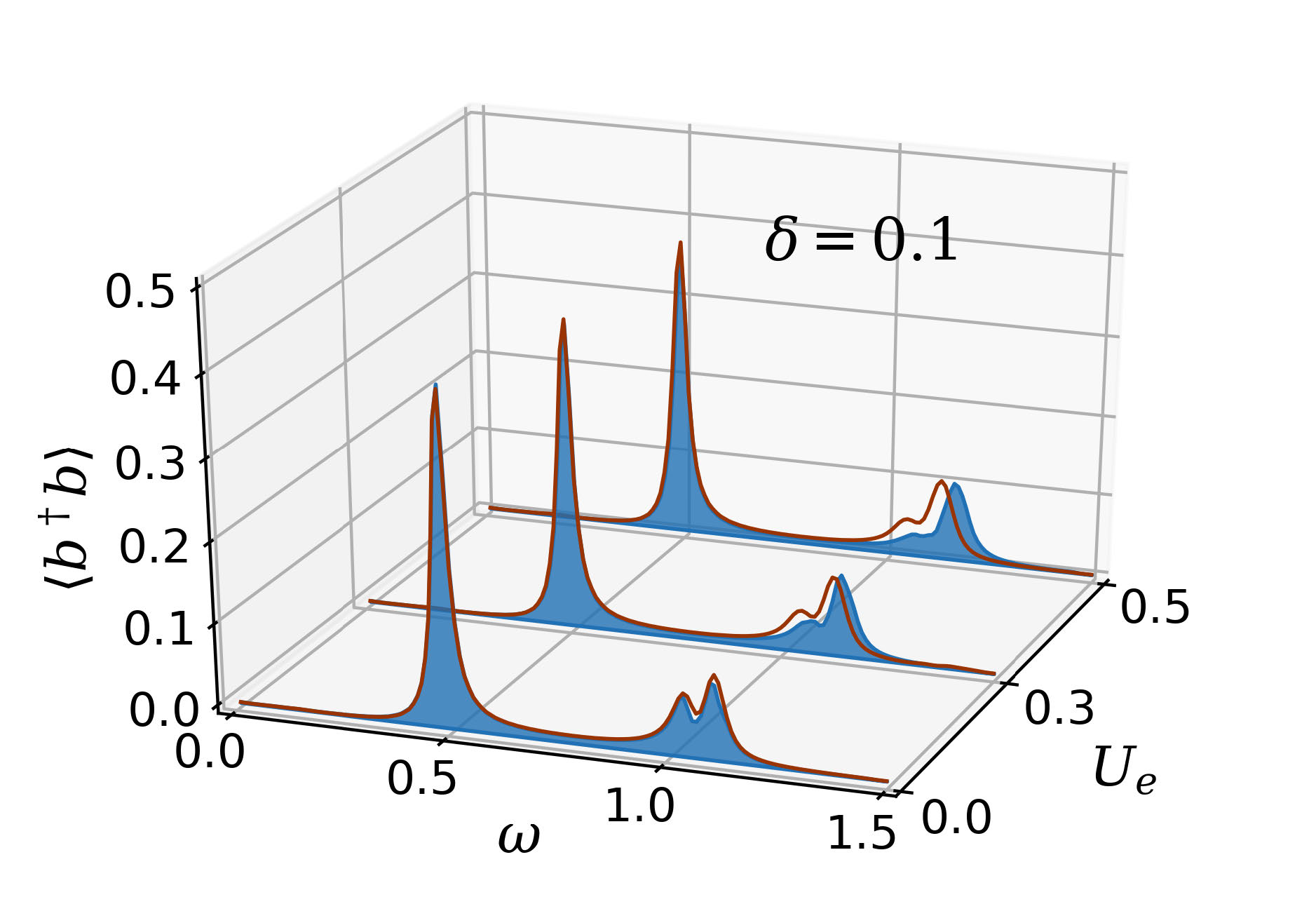}};
  \node at (12.0,-2.50) {\includegraphics[width=0.34\columnwidth]{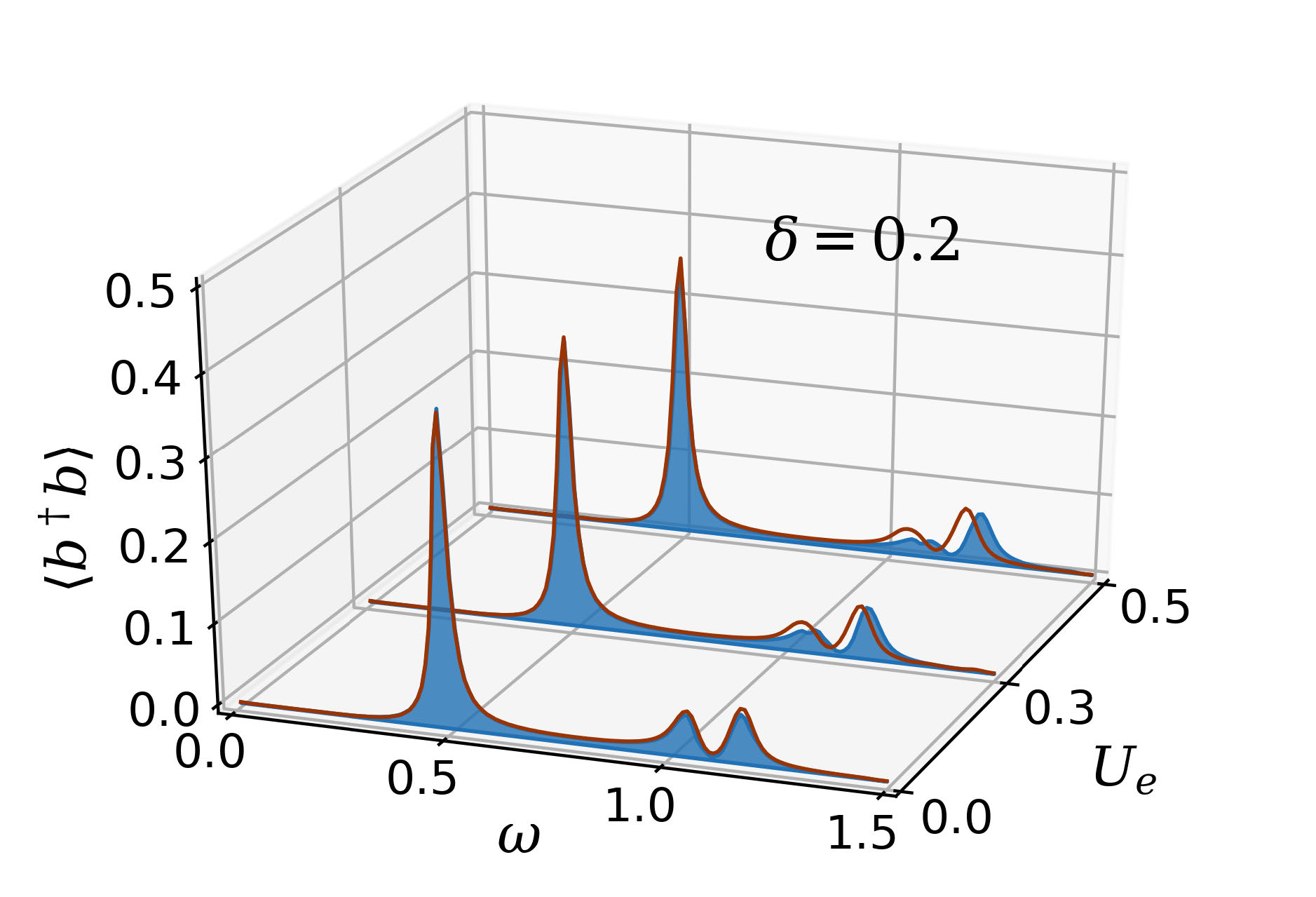}};
  \node[black] at (-1.5, 3.0) {$\bm{(a)}$};
  \node[black] at ( 4.5, 3.0) {$\bm{(b)}$};
  \node[black] at (10.6, 3.0) {$\bm{(c)}$};
  \node[black] at (-1.5,-1.50) {$\bm{(d)}$};
  \node[black] at ( 4.5,-1.50) {$\bm{(e)}$};
  \node[black] at (10.6,-1.50) {$\bm{(f)}$};
  \node[black] at (5.5, 4.0) {\Large{\bm{$L=3$}}};
  \node[black] at (5.5,-0.50) {\Large{\bm{$L=7$}}};
 \end{tikzpicture}
 \caption{SHG spectrum from ED (brown curves) and NEGF-GKBA (solid filled blue curves) for $L=3$ (upper panel) and $L=7$
   (lower panel). The results are for disorder strengths $\delta=0.0 \,(\bm{a}, \bm{d}),\,\delta=0.1\,(\bm{b}, \bm{e})$
   and $\delta= 0.2\,(\bm{c}, \bm{f})$, and interaction $U_e=0.0,0.3$ and $ 0.5$. All curves are shown at $t_e=250$, the
   end point of the time-evolution interval.}
 \label{fig1}
\end{figure*}
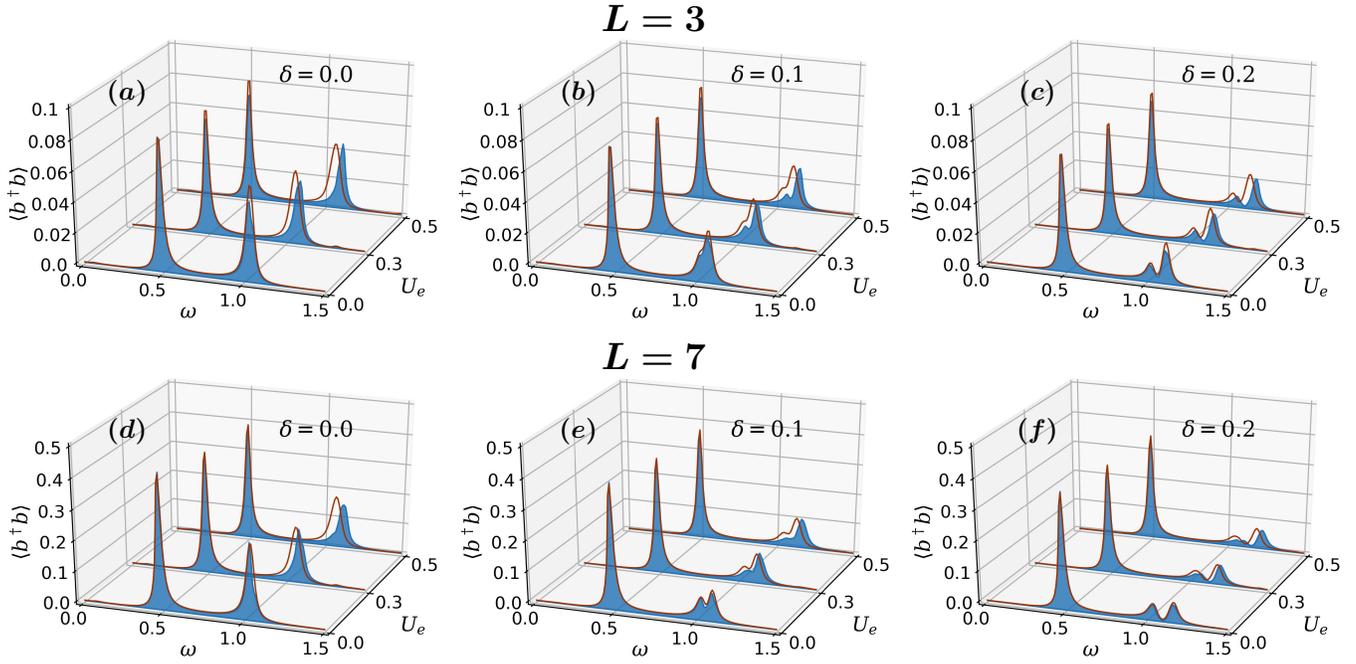

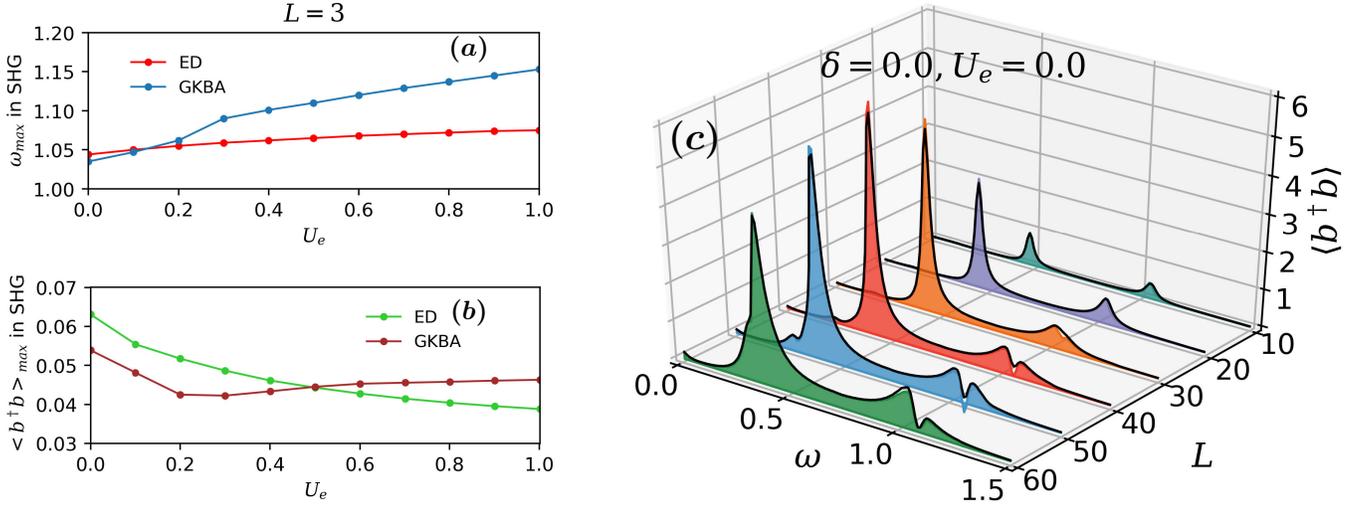
\begin{figure}
	\begin{tikzpicture}
	\node at (0.0, 3.5) {\includegraphics[width=0.4\columnwidth]{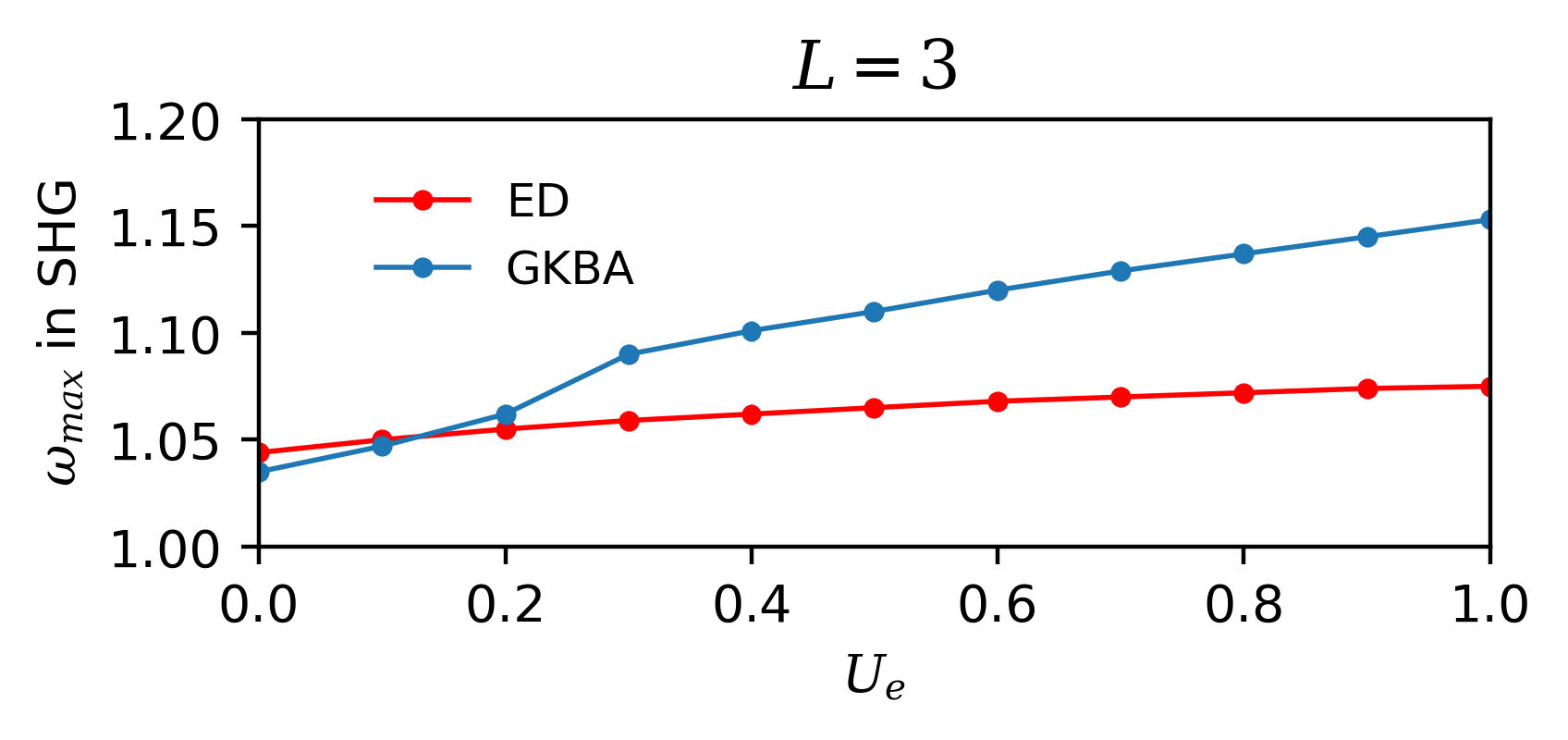}};
	\node at (0.0, 0.0) {\includegraphics[width=0.4\columnwidth]{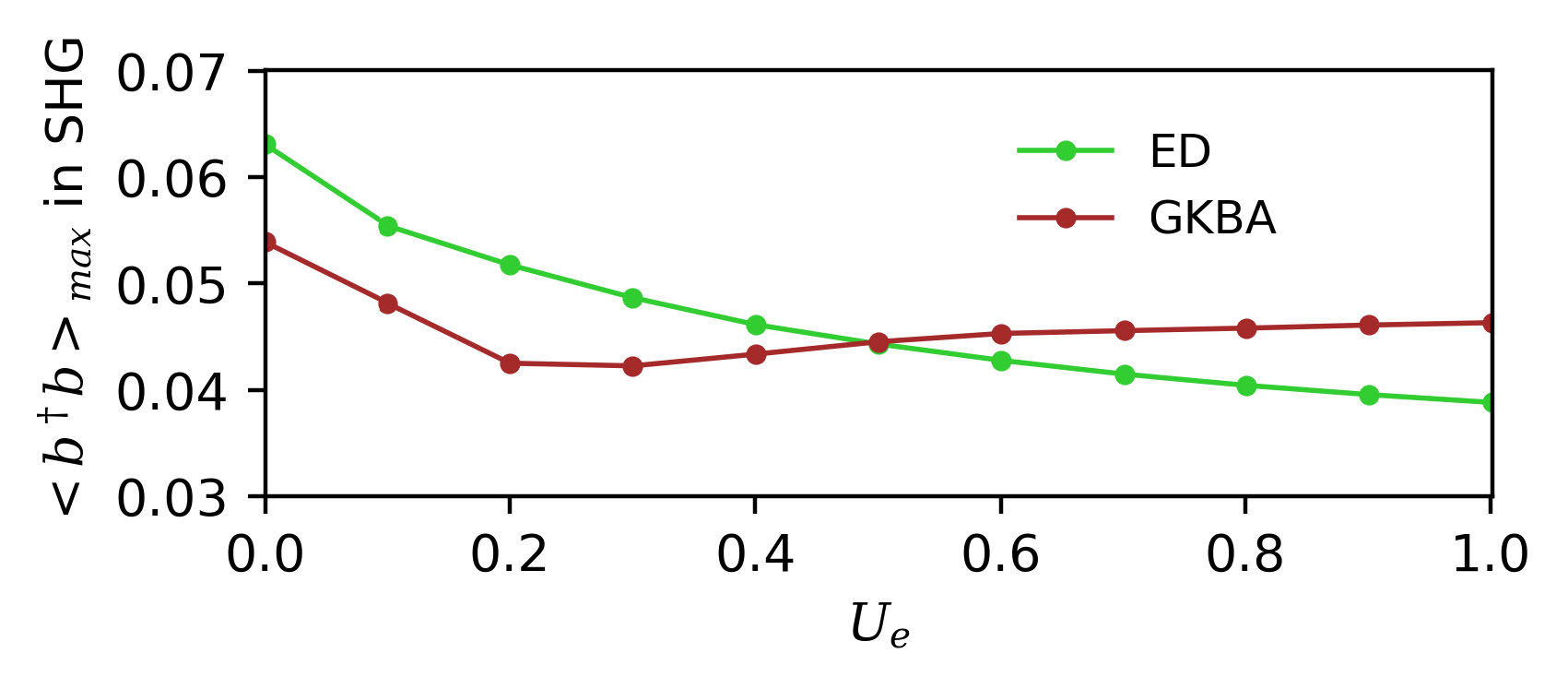}};
	\node at (9.0, 2.0) {\includegraphics[width=0.56\columnwidth]{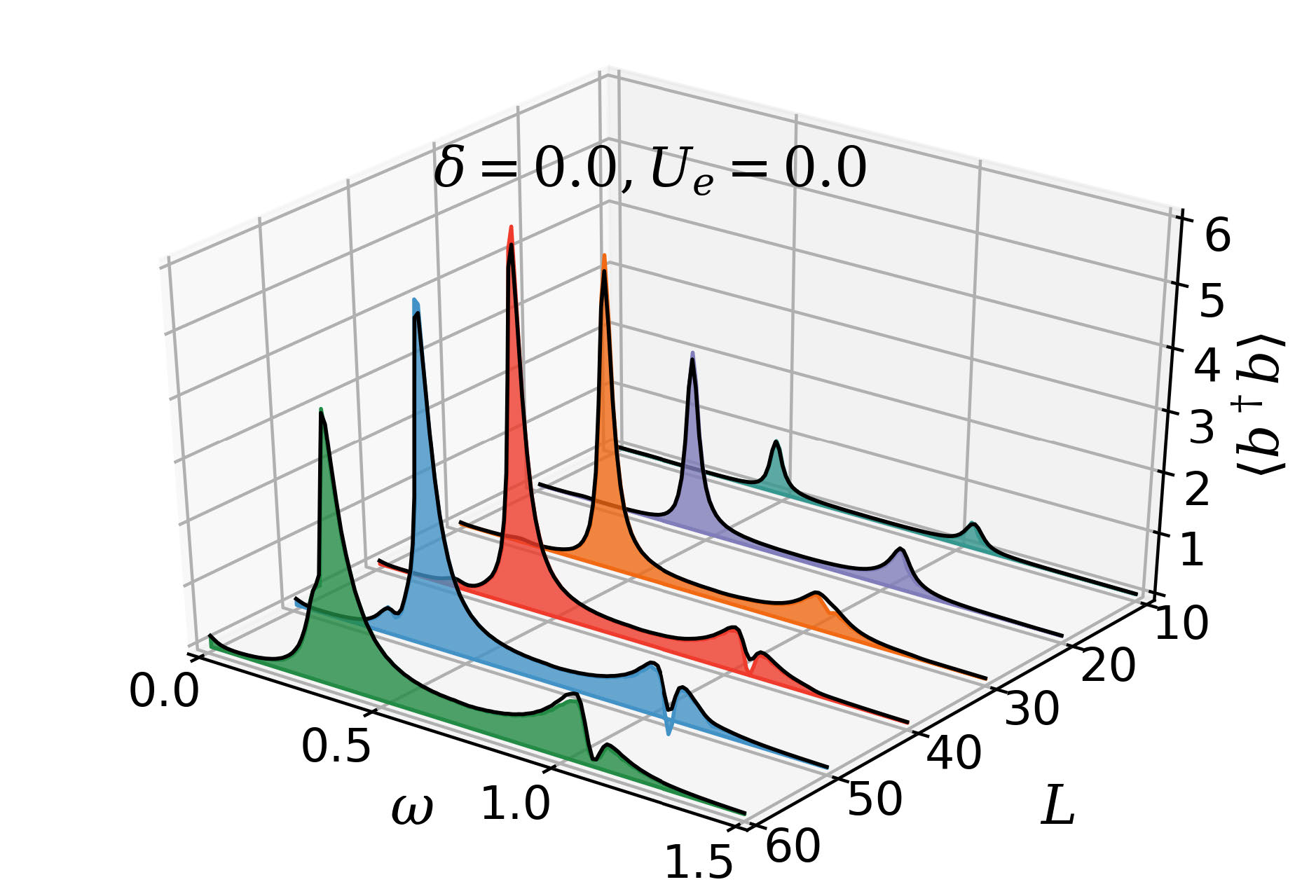}};
	\node[black] at (2.5, 4.50) {$\bm{(a)}$};
    \node[black] at (2.5, 1.0) {$\bm{(b)}$};
    \node[black] at (5.5, 3.3) {\Large${\bm{(c)}}$};
	\end{tikzpicture}

	\caption{ED and NEGF-GKBA results at $L=3$ and $\delta=0.0$ for the frequency $\omega_{max}$ (panel $\bm{(a)}$)
          and photon occupation $\langle\hat b^\dagger \hat b\rangle$ (panel $\bm{(b)}$) in correspondence of the maximum
          intensity of the SHG spectrum. Panel ${\bm{(c)}}$: ED (black empty curve) vs GKBA (solid filled curves)
          results for Dicke systems at $\delta=U_e=0.0$ and $L=10,20,30,40,50$ and $60$.In all panels, the ED
          calculations are performed with the spin formalism, and all curves are calculated at time $t_e=250$.}
	\label{fig2}
\end{figure}

To benchmark the NEGF-GKBA approach, we have considered ED calculations for a smaller number of TLS, and compared these
with GKBA results. In most cases, both disorder (in the sense specified in Eq.~(\ref{periodis})) and e-e interactions
are included. The comparisons are shown in \textbf{Figure \ref{fig1}}, where it is manifest that overall ED and GKBA
results agree very well. Good agreement is observed also in the presence of disorder. However, with e-e interactions
included, the SHG peak from GKBA is shifted slightly towards the higher frequencies compared to ED.
To characterize these discrepancies in more detail, we singled out the case of three TLS (i.e., $L=3$) for several
values of $U_e$ and, to establish trends, we focused on the behavior of the maximum intensity in the SHG peak
$\langle\hat b^\dagger \hat b\rangle_{max}$ in relation of the corresponding emitted frequency $\omega_{max}$. This was
done for both ED and GKBA, and the inherent results are shown in \textbf{Figure \ref{fig2}}$\bm{(a)}$ and
\textbf{\ref{fig2}}$\bm{(b)}$. In ED, at larger $U_e$, $\omega_{max}$ increases only slightly (panel $\bm{(a)}$),
i.e. the SHG peak appears to be not very sensitive to $U_e$ (at least in this interaction range). However, with GKBA, a
larger shift of the SHG peak occurs with the increase of $U_e$.

Coming now to the SHG spectral intensity, in panel $\bm{(b)}$ we observe that, in the ED calculations, increasing $U_e$
reduces the intensity of the SHG peak, whereas, in GKBA, the trend of decreasing intensity is observed only for lower
$U_e$; indeed, for higher $U_e$ the intensity almost remains constant. It is beyond the scope of this short
contribution to address in a quantitative way the reasons for these small but clearly discernible discrepancies. Still,
it can be of interest to provide some additional considerations. A first factor to take into account is that, together
with photon dressing effects, the system's energy levels undergo { multiple} level crossing as a function of the
interaction $U_e$ ( in particular, the level crossing at $U_e=0$ represents an essential complication, which is not easy
to account in perturbation theory. The second aspect to consider is that, in any case, an HF treatment of e-e
interactions can be inadequate, as shown next with simple qualitative argument in the large-$U_e$ limit.

Specifically, in our model (where interactions only occur among electrons in the upper levels of the TLSs), the ground
state within an HF description is a state with, say, a fraction $\delta$ of electrons on the upper level.  This fraction
can be estimated by balancing the energy between the electrons and the $\omega_a$-type photons: $U_e\delta^2 \approx {
  \omega_a\beta^2 \rightarrow \delta \approx \sqrt{\omega_a\beta^2/U_e}}$. Excited states in the HF theory are rather
simple and similar to the exact ones (in our case we basically have a spin-flip transition of an electron from the lower
to the upper state in the TLS). The energy of such a transition, in addition to the noninteracting part equal to 1,
contains a mean field contribution $U_e [(1-\delta)^2-\delta^2]= U_e(1-2\delta)={ U_e[1-2\sqrt{\omega_a\beta^2/U_e}]}$.
This yields in total for the fluorescence $\omega \approx 1+U_e[1-2\sqrt{\omega_a\beta^2/U_e}]$, in contrast to
$\omega \simeq 1$ in the exact theory.

Even so, it is important to restate here that the overall performance of the NEGF-GKBA is very good, together with the
observation (stimulated by preliminary estimates, not shown here) that the differences observed most likely are due to
the shortcomings of the HF treatment for the e-e interactions, and not of the photon-electron interactions.

\subsection{Spin representation in ED for large Dicke systems}
The comparisons performed so far, while successful for NEGF-GKBA, concerned rather small systems.  However, a main aim
of this work is to explore the scope of NEGF-GKBA for large and interacting and/or disordered Dicke systems, in general
not accessible to ED.  Yet, at least in the absence of disorder and e-e interactions, it is still possible to use ED as
reference for $L$ considerably larger than in the previous section.  To make this explicit, we consider again the spin
operators representation so that (dropping again { the irrelevant} total energy shift in {\color{black} $\hat H_S$}) the
Hamiltonian in Eq.~(\ref{sys_Ham}) becomes
\begin{equation}
  \hat{\tilde{H}}_S={\Delta} \hat{S}^z+\omega_a \hat a^\dagger \hat a +\omega \hat b^\dagger \hat b
  +2\left[ g_a(\hat a^\dagger+\hat a)+g_b(t)(\hat b^\dagger+\hat b)\right]\hat{S}^x,
\end{equation}
where $\hat{S}^{x,z}=\sum_{\bm{i}=1}^L \hat s_{ \bm i}^{x,z}$.

In this form, $\hat{\tilde{H}}_S$ manifestly commutes with the total spin $\hat{S}^2$ operator, and thus the
photon-induced transitions conserve the total spin of the system. Hence, starting from an initial state of definite
total spin (e.g., with the electron in each TLS in the lower level, corresponding to $S=L/2$ and $S_z=-L/2$), the
dynamics of the non-disordered and non-interacting electron system induced by the photon fields can be represented in ED
within a Hilbert space of $L+1$ (instead of $2^L$) basis states. This makes possible to consider ED benchmarks (for
$U_e=\delta=0$) when $L$ is large, and thus have a further comparison with NEGF-GKBA.  The comparison is shown in
\textbf{Figure \ref{fig2}}$\bm{(c)}$, with a black empty curve representing ED results, and the coloured solid curve
corresponds to GKBA results. We can observe that ED and GKBA results are in excellent agreement in the considered
parameter range. Commenting on the physical information in these results, we note that until $L=40$, the intensity of
the spectra increases with $L$, which is expected. However, for $L>40$, even though there is a small increase in the
intensity of SHG with $L$, the intensity of the main peak (peak close to $\omega=0.5$) decreases with $L$.

\begin{figure}
\centering
	\begin{tikzpicture}
	\node at (-2.3, 2.3) {\includegraphics[width=0.4\columnwidth]{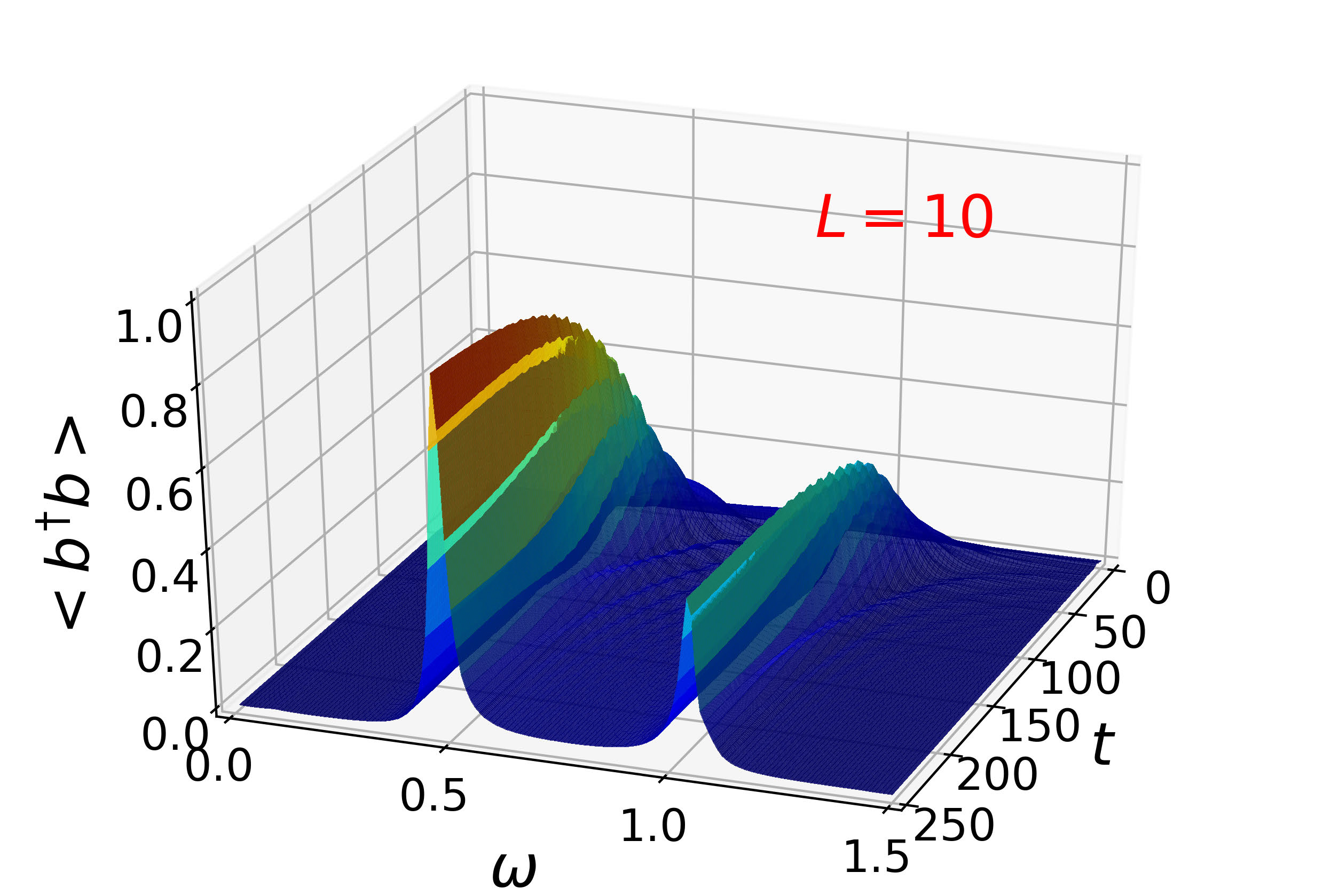}};
	\node at (5.1,2.3){\includegraphics[width=0.4\columnwidth]{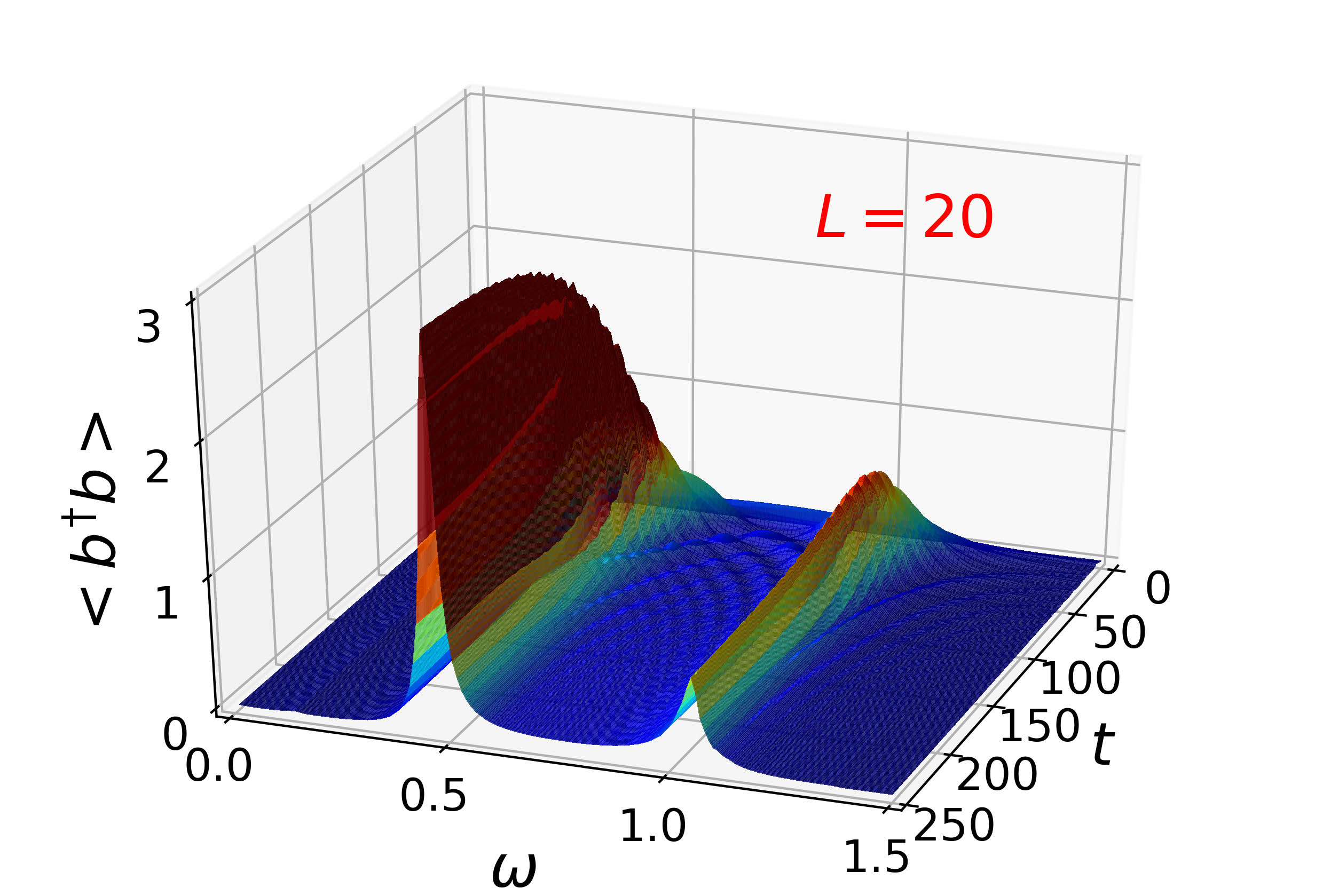}};
	\node at (-2.3, -2.90) {\includegraphics[width=0.4\columnwidth]{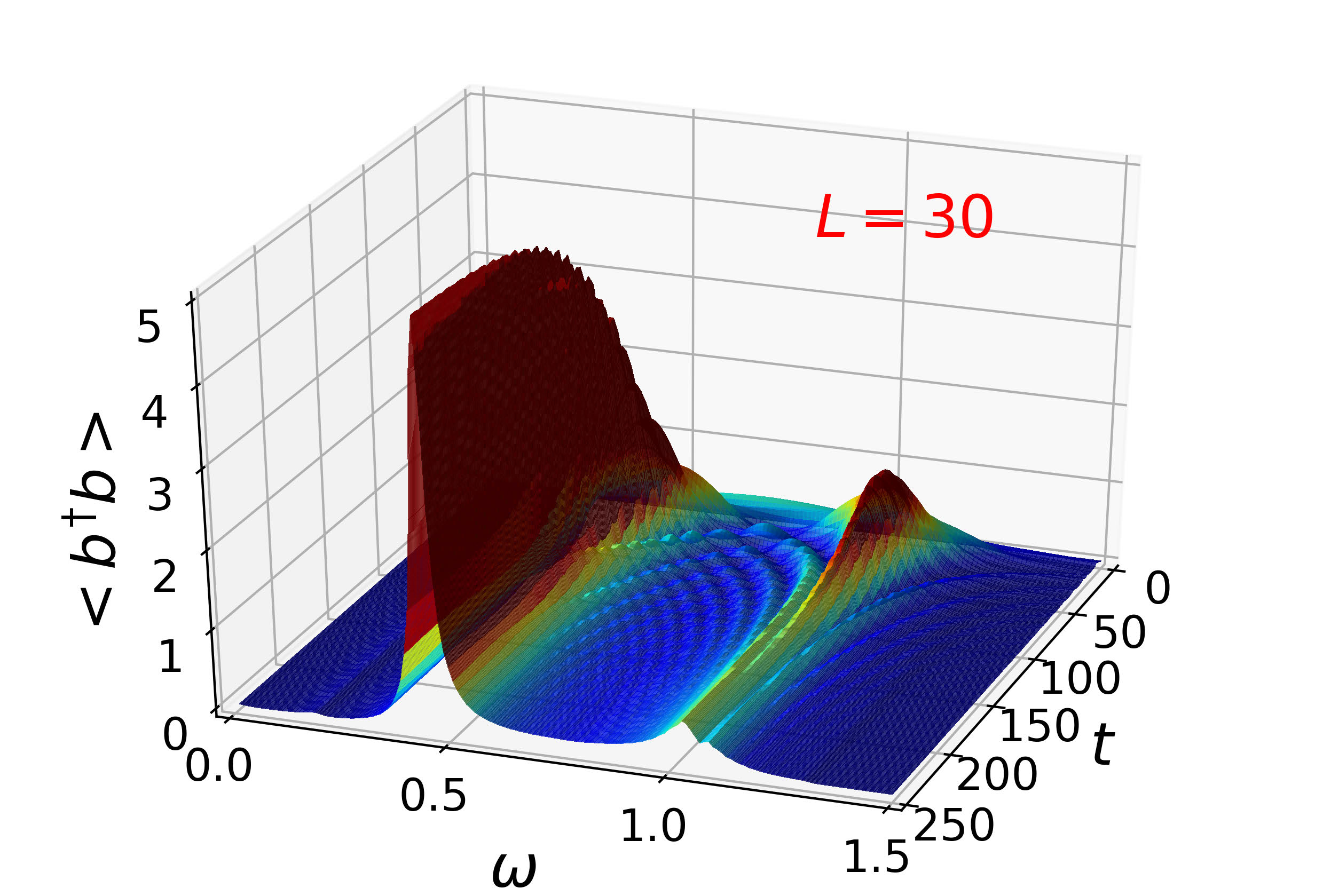}};
	\node at (5.1, -2.90) {\includegraphics[width=0.4\columnwidth]{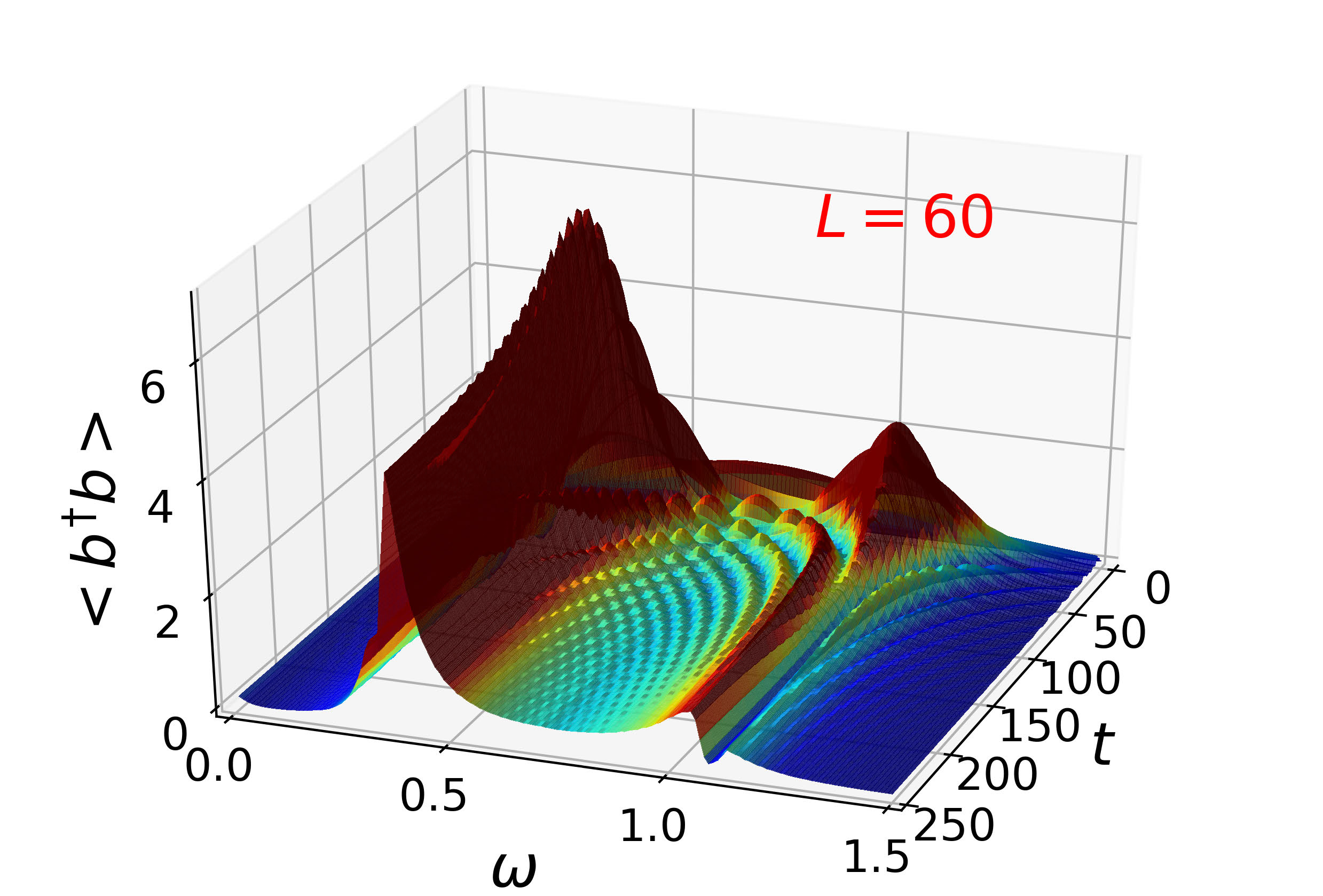}};
	
	    \node[black] at (-4.40,3.2) {\large$\bm{(a)}$};
     	\node[black] at (3.0,3.2) {\large$\bm{(b)}$};
     	\node[black] at (-4.40,-2.) {\large$\bm{(c)}$};
     	\node[black] at (3.0,-2.) {\large$\bm{(d)}$};
	\end{tikzpicture}
	\caption{ Time dependent NEGF-GKBA spectra for $\delta=U_e=0.0$ and $L=10,20,30,60$.  The parameters used are
          $\epsilon_e-\epsilon_g \equiv \Delta=1.0, g_a=0.03, g'=0.01, \hat n_a=\langle \hat a^\dagger \hat a
          \rangle=\beta^2=9,\Gamma=0.02$, and $\omega_a=0.5$ All curves are shown at $t_e$=250, the end point of the
          time-evolution interval.}
	\label{fig3}
\end{figure}
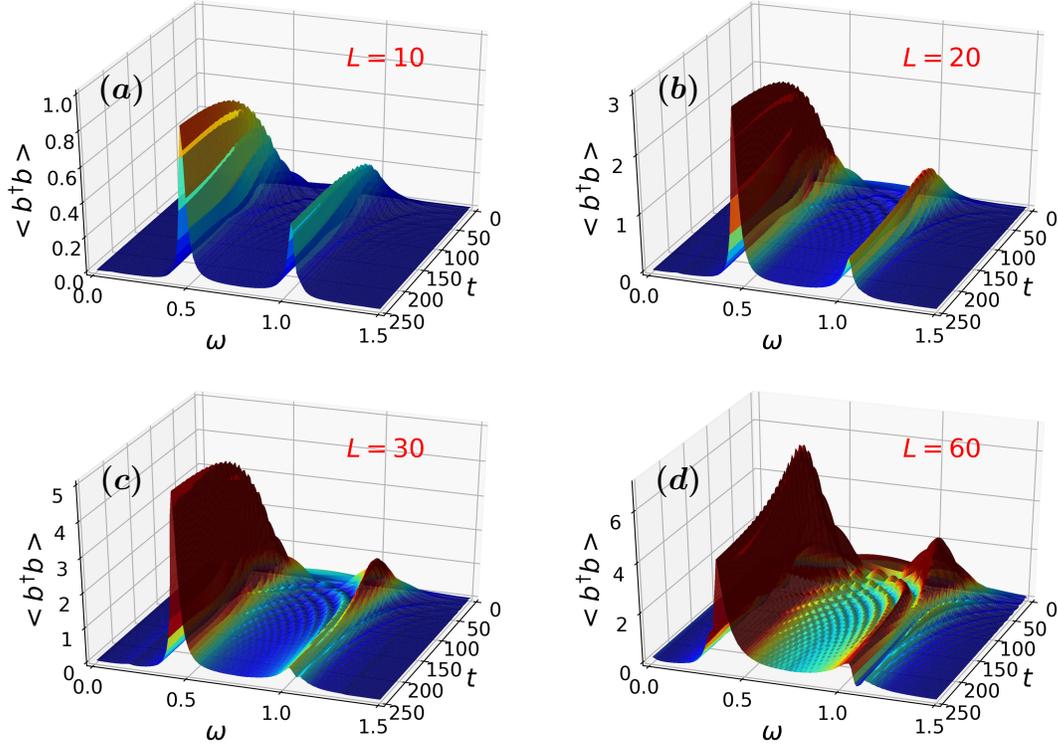

Figure \ref{fig2}$\bm{(c)}$ shows the time evolved plots obtained at a final time $t_e=250$.  It can be useful to
consider the actual development in time of such spectra, that is reported in \textbf{Figure \ref{fig3}}. During the time
evolution, and for $L< 20$, the intensity of the spectra increases initially but then remains almost constant. However,
for larger Dicke systems ($L>20$), the trend is not monotonic: even though we initially observe a rise in the intensity
in time, at some point the signal will start to drop, which is consistent with the results of {Figure
  \ref{fig2}}$\bm{(c)}$ in the long time limit. In summary, for the parameters considered, a steady increase in the
intensity of the SHG spectra on increasing $L$ was not observed.

\section{Disorder and interaction}\label{disordinato}

\begin{figure}
	\begin{tikzpicture}
	\node at (0.0, 2.3) {\includegraphics[width=0.5\columnwidth]{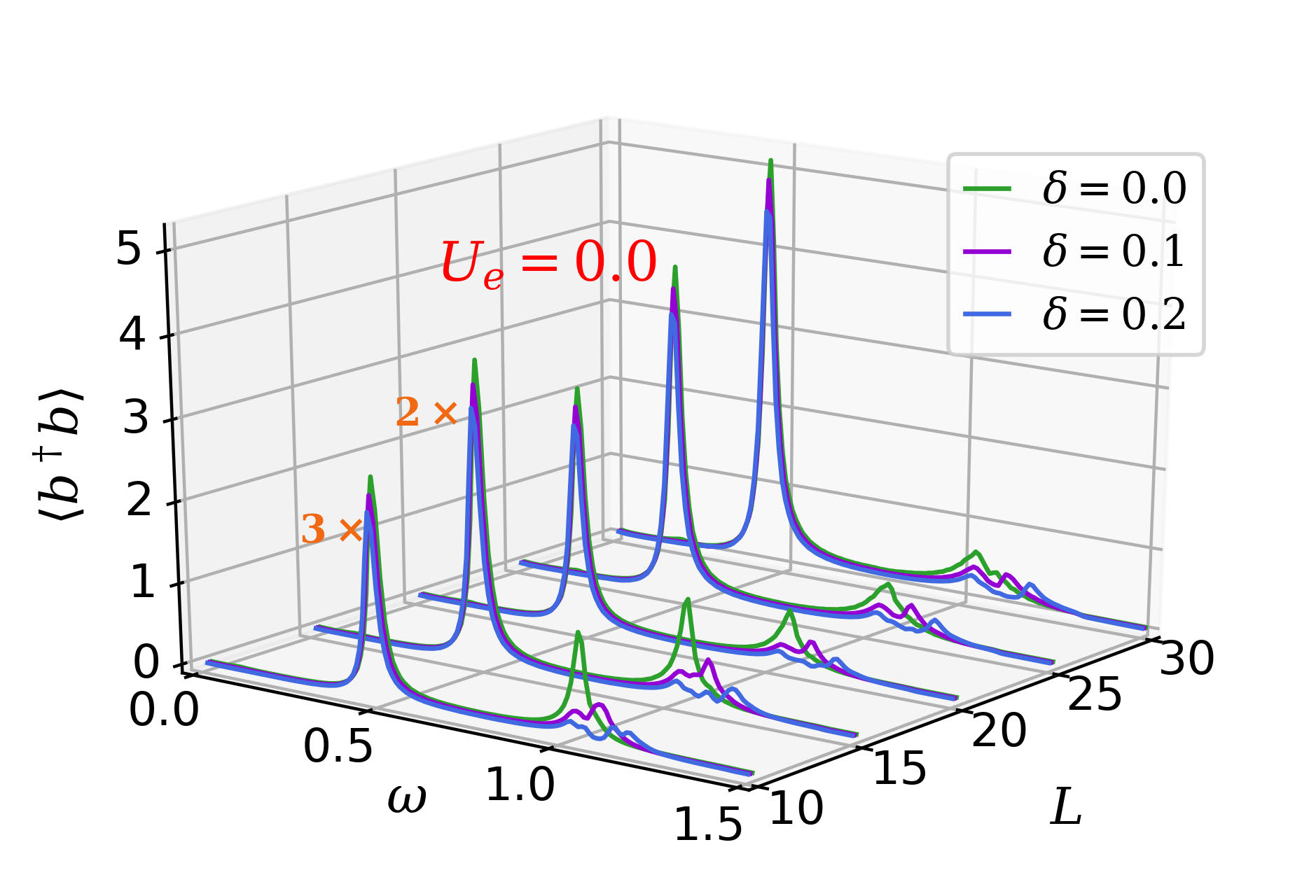}};
	\node at (9.2, 2.3) {\includegraphics[width=0.5\columnwidth]{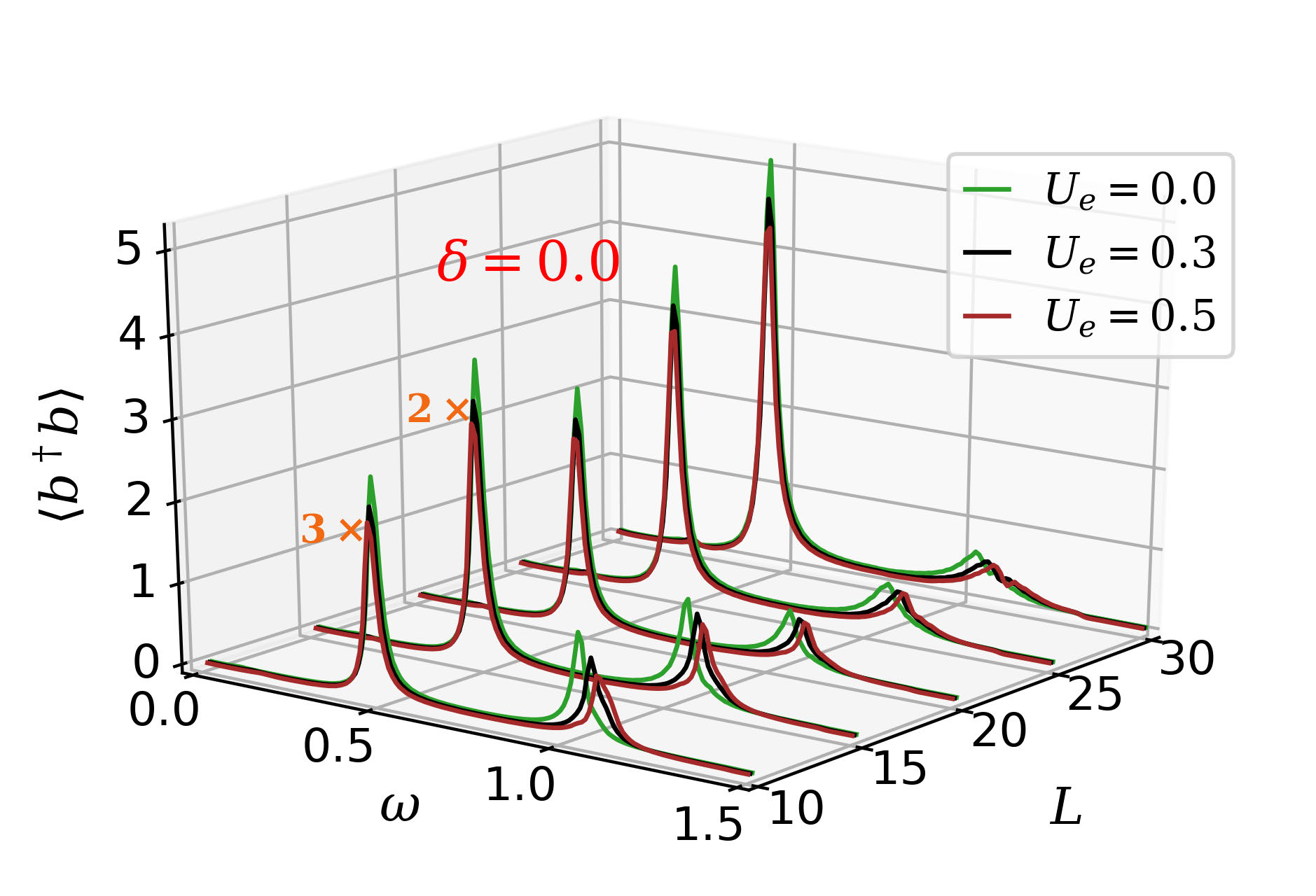}};
	\node[black] at (-2.80,3.4) {\large$\bm{(a)}$};
    \node[black] at (6.40,3.4) {\large$\bm{(b)}$};
	\end{tikzpicture}
	\caption{Panel $\bm{(a)}$: spectra in the presence of disorder, with $\delta=0.0,0.1,0.2$ and $U_e=0.0$. Panel
          $\bm{(b)}$: spectra in the presence of interactions, with $U_e=0.0,0.3,0.5$ and $\delta=0.0$. The systems
          considered have $L=10,15,20,25,30$. The spectra for $L=10$ are scaled by `$3$' and $L=15$ by `$2$' for visual
          clarity. All spectra were obtained at the time $t_e=250$.}
	\label{fig4}
\end{figure}

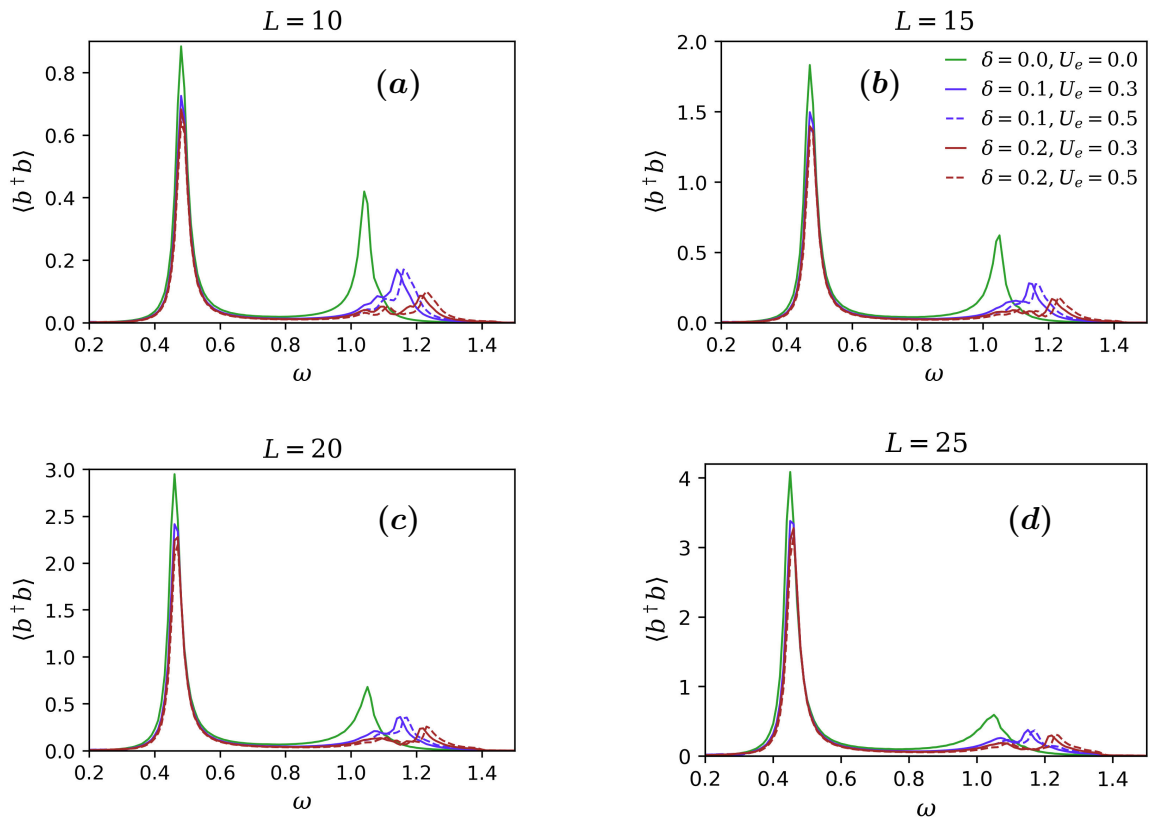
\begin{figure}
   \centering
	\begin{tikzpicture}
	\node at (-2.3, 2.3) {\includegraphics[width=0.37\columnwidth]{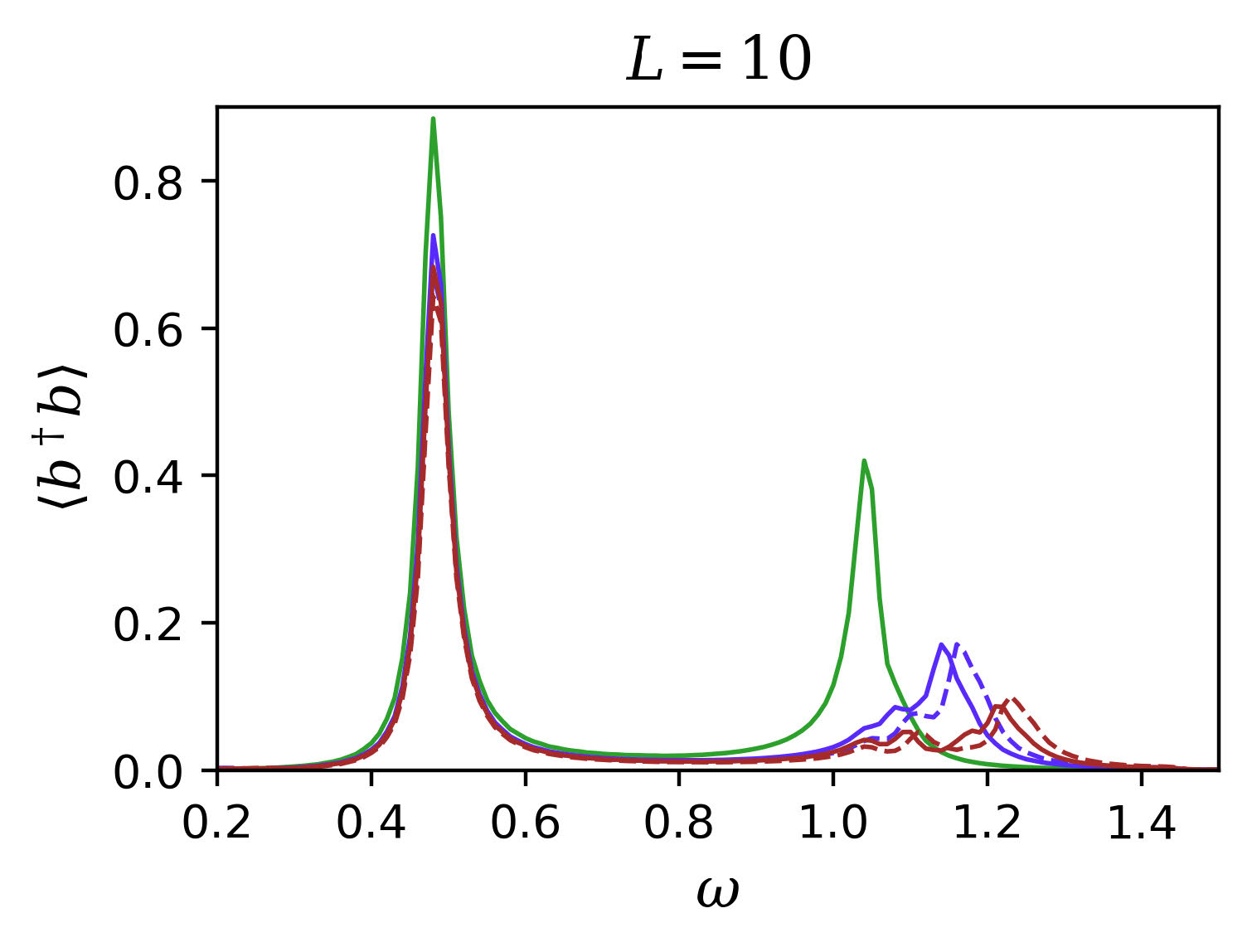}};
	\node at (6.1,2.3){\includegraphics[width=0.37\columnwidth]{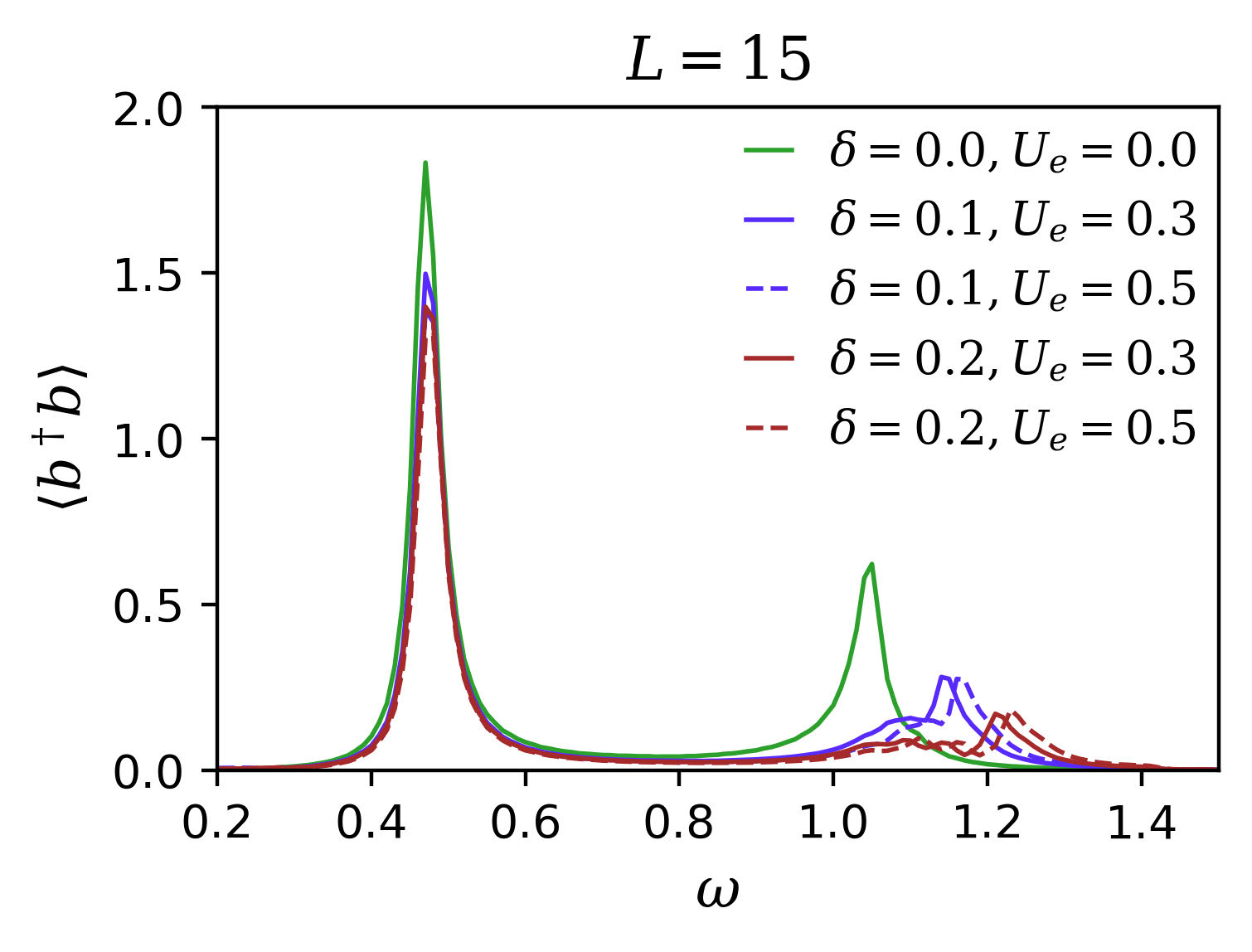}};
	\node at (-2.3, -3.40) {\includegraphics[width=0.37\columnwidth]{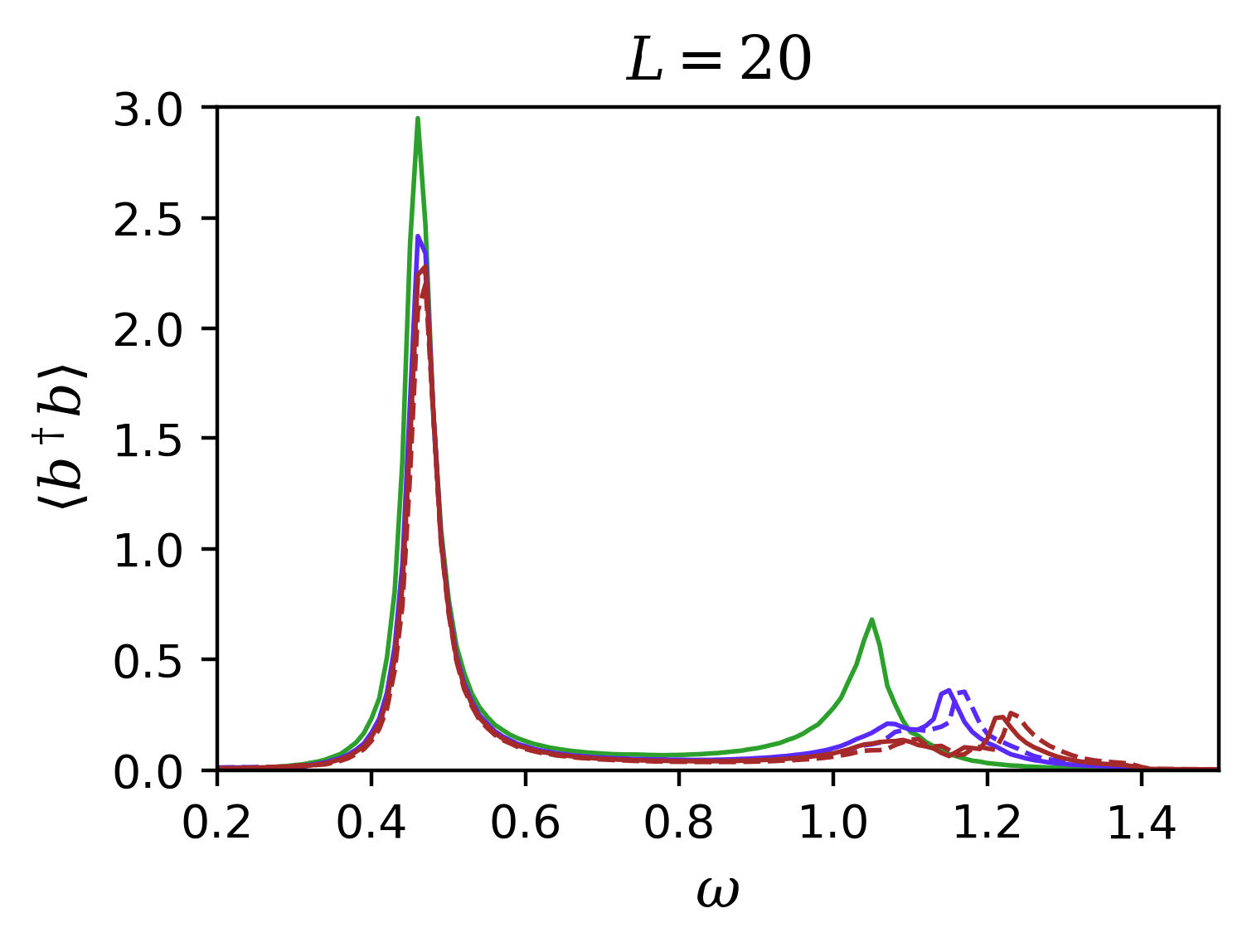}};
	\node at (6.1, -3.40) {\includegraphics[width=0.37\columnwidth]{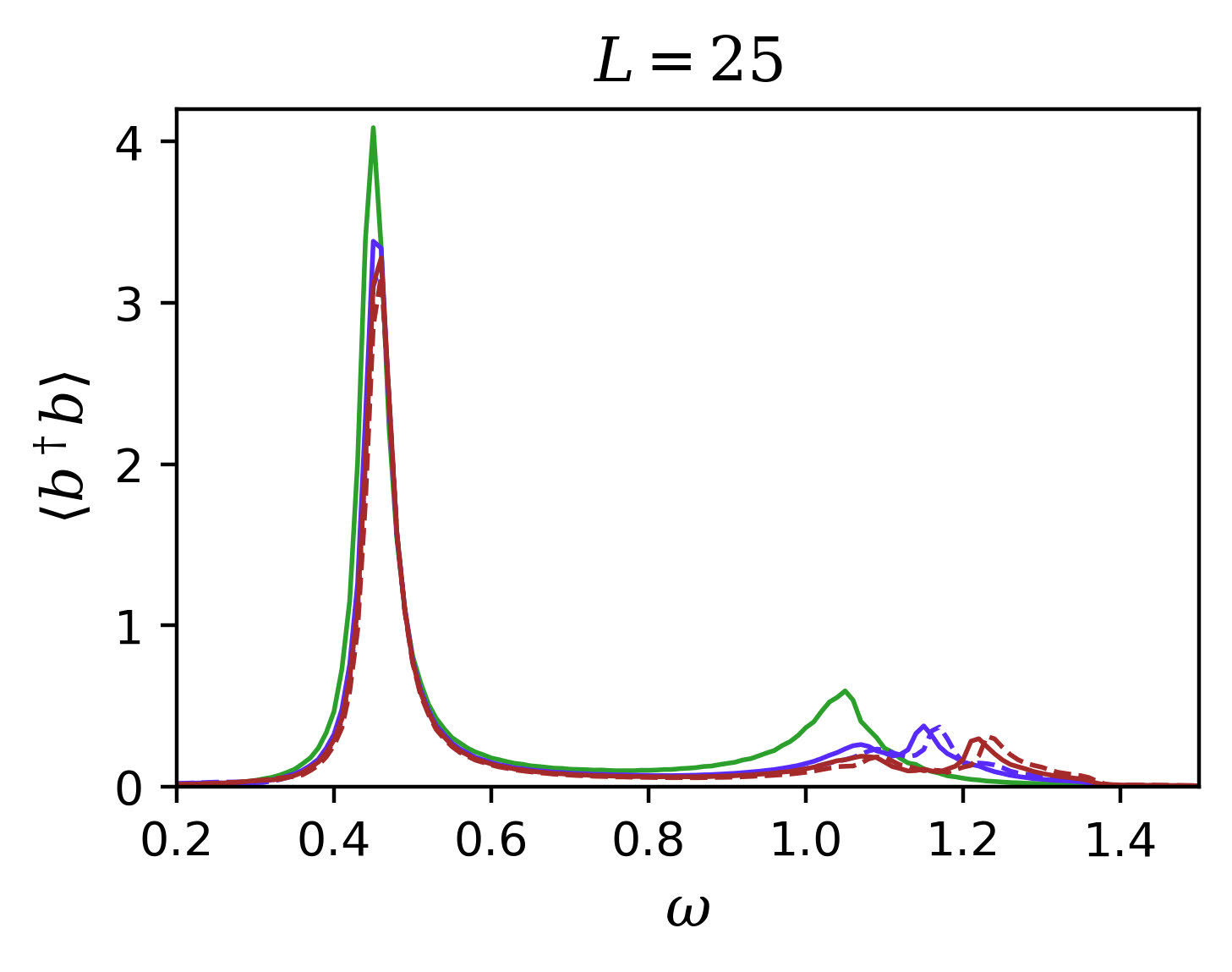}};
	
	\node[black] at (-0.50,3.8) {\large$\bm{(a)}$};
     	\node[black] at (5.90,3.8) {\large$\bm{(b)}$};
     	\node[black] at (-0.50,-2.0) {\large$\bm{(c)}$};
     	\node[black] at (7.90,-2.0) {\large$\bm{(d)}$};
	\end{tikzpicture}
	\caption{ Fluorescent spectra obtained with NEGF-GKBA for $L=10,15,20,25$, disorder strength $\delta=0.0$ and
          interaction strength $U_e=0.0$. The spectra were calculated at the time $t_e=250$.}
	\label{fig5}
\end{figure}

We now consider the role of disorder and electron interactions on SHG for the larger Dicke systems, and only within a
NEGF-GKBA treatment. Disorder and interactions are included either one at the time (\textbf{Figure \ref{fig4}}), to
disentangle their effect, or together (\textbf{Figure \ref{fig5}}), to see if they are mutually competing or synergic in
affecting the spectrum.

SHG spectra for $U_e=0$ and different disorder strengths (\textbf{Figure \ref{fig4}}$\bm{(a)}$) show that at larger
$\delta$ the SHG spectral will diminish, along with some peak splitting.  Increasing $\delta$ will set the resonance
frequency ($\omega_{res}$) of several TLS: in the Dicke system away from $\Delta$. Hence the SHG peak intensity
decreases. Also, the different TLS of the disordered system will have different resonance frequencies; hence, there are
multiple peaks around the nominal SHG frequency. Conversely, in \textbf{Figure \ref{fig4}}$\bm{(b)}$ there is no
disorder ($\delta=0.0$), and one can observe the sole effect of electron interaction. On increasing the electron
interaction, the intensity of the SHG peak reduces again. But the intensity reduction observed with the electron
interaction is less compared to the disorder induced reduction.  Finally, we also considered calculations with both
disorder and interactions present (\textbf{Figure \ref{fig5}}). We did not observe any mutual compensation/competition
of disorder and interactions effects; rather, the intensity of the SHG peak got reduced even further.

\section{Cavity leakage}\label{perdite}
In reality, photons inside the cavity will escape from the cavity after some time. In the case of a bad cavity, the
photons will escape more quickly. We now mimic the case of a bad cavity by introducing cavity leakage via coupling the
photon fields to a bath of classical harmonic oscillators. This method is inspired from the physics of the
Caldeira–Leggett model \cite{Caldeira1983,Venkataraman2014}. Hence, the system Hamiltonian $\hat H_S$ in Eq.~\ref{sys_Ham}
will be augmented with the two extra terms $\hat H_{ph-bath}$ and $\hat H_{bath}$ \cite{Scipost}. The interaction between the
photon field and the bath is represented by
\begin{equation}
 \hat H_{ph-bath}=\sum_{k}^{N_{bath}} C_k\left[{q_{a,k}}(t)\big(\hat a^\dagger + \hat a\big) +{q}_{b,k}(t)\big(\hat
   b^\dagger+\hat b\big)\right],
\end{equation}
where ${q_{a,k}}$ and ${{q}_{b,k}}$ are the coordinates of the oscillators, the coupling strength is $C_k$ and $C_k=A({
  \Delta_B} k)^a$.

The Hamiltonian {\color{black} function} corresponding to the baths of the classical oscillators is
\begin{equation}
 H_{bath}=\sum_{k} ^{N_{bath}}\Big(\frac{{p_{a,k}^2}+{p_{b,k}^2}}{2m_k}+\frac{1}{2}\omega_k^2m_k \big[{q_{a,k}^2}+{{q}_{b,k}^2}\big]\Big),
\end{equation}
where $p_{a,k} \,(p_{b,k})$ represents the momentum of the $k$th oscillator coupled to the cavity (fluorescent) field,
and $\omega_k={ \Delta_B} k$ corresponds to the frequency of the oscillator.

In terms of the operator $\hat \phi_\mu$ defined in Sect.~\ref{negf_method}, one can express the quantum-classical
Hamiltonian as $\hat H_{ph-bath}=\sum_\mu F_\mu(t)\hat \phi_\mu$ where, explicitly, $\{F_\mu\}\equiv (F_{1,1}, F_{1,2},
F_{2,1}, F_{2,2})$, with $F_{1,1}=\sqrt 2\sum_k^{N_{bath}}C_k{q_{a,k}}(t)$, $F_{2,1}=\sqrt 2\sum_k^{N_{bath}}C_k{
  q_{b,k}}(t)$ and $F_{1,2}=F_{2,2}=0$.

\begin{figure*}
 \begin{tikzpicture}
  \node at ( 0.0, 2.0) {\includegraphics[width=0.33\columnwidth]{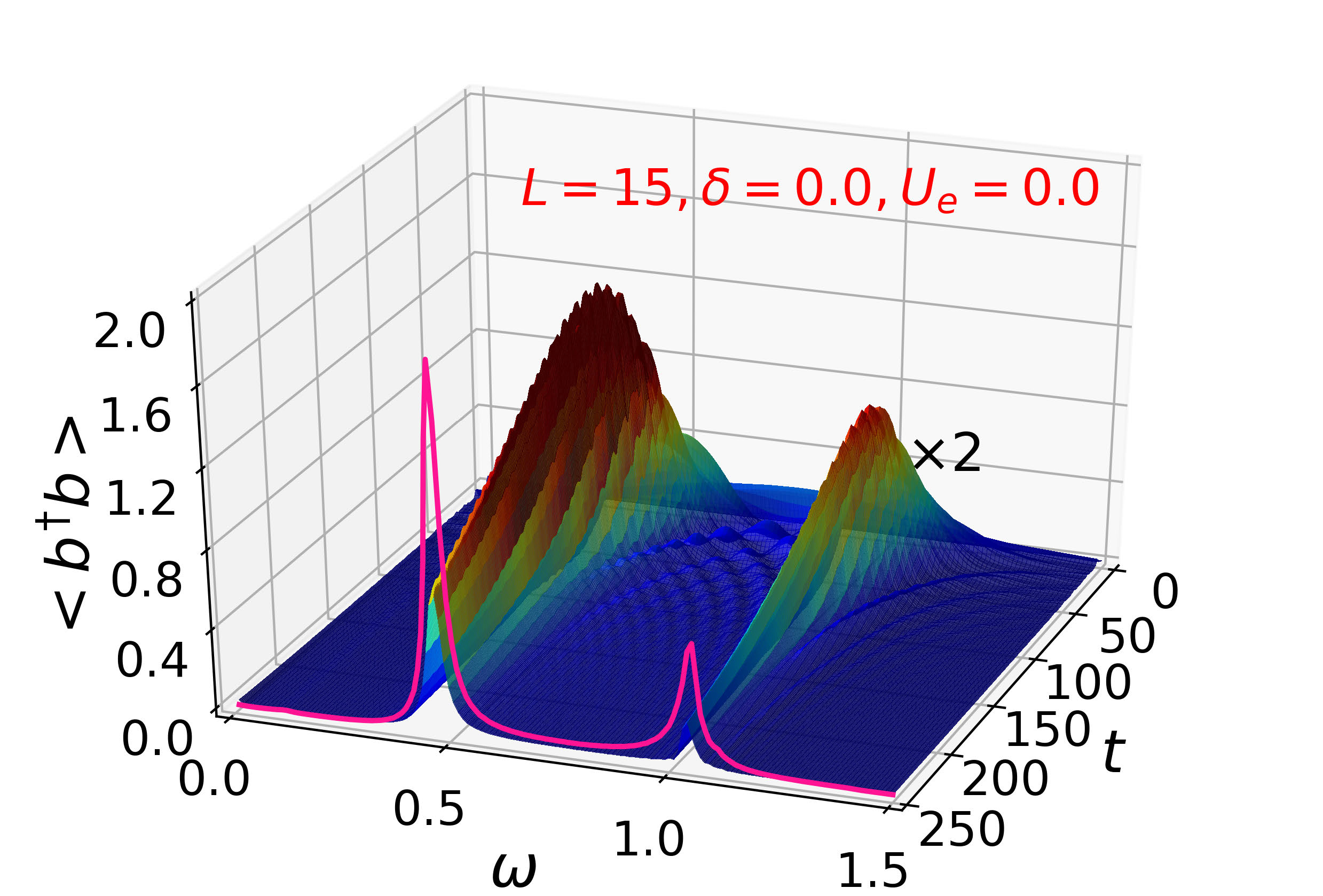}}; 
  \node at ( 6.0, 2.0) {\includegraphics[width=0.33\columnwidth]{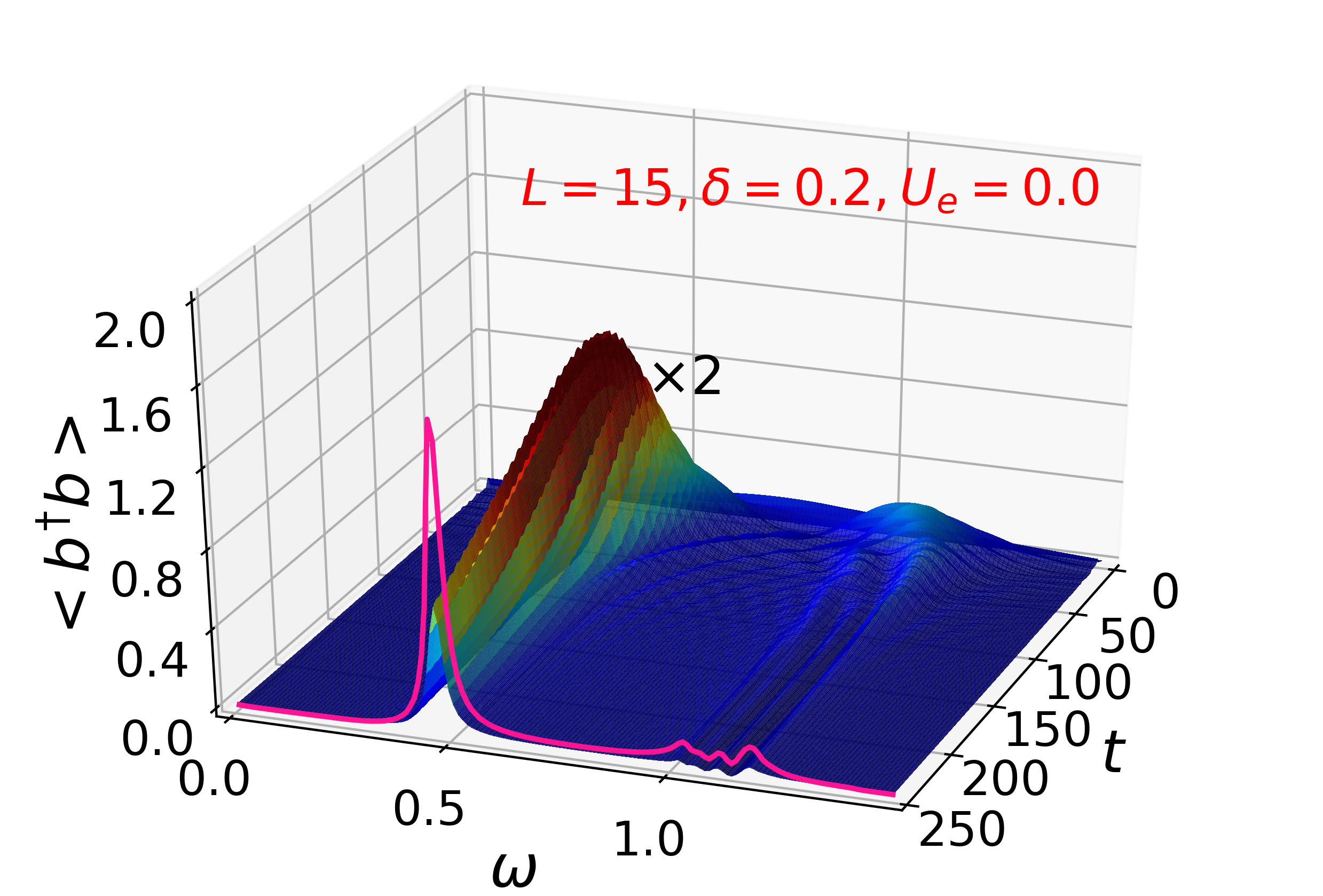}};
  \node at (12.0, 2.0) {\includegraphics[width=0.33\columnwidth]{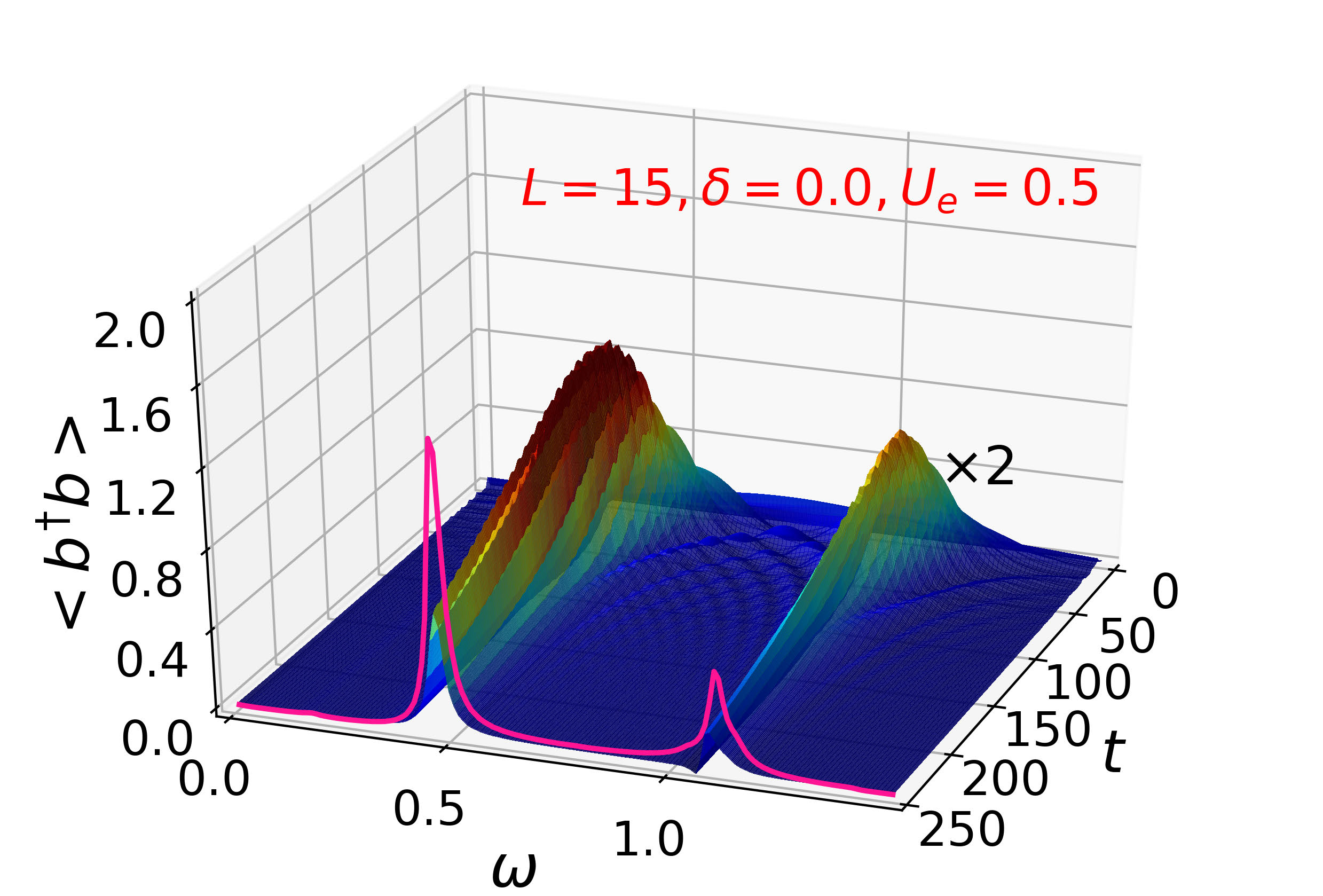}};
  \node at ( 0.0,-2.90) {\includegraphics[width=0.33\columnwidth]{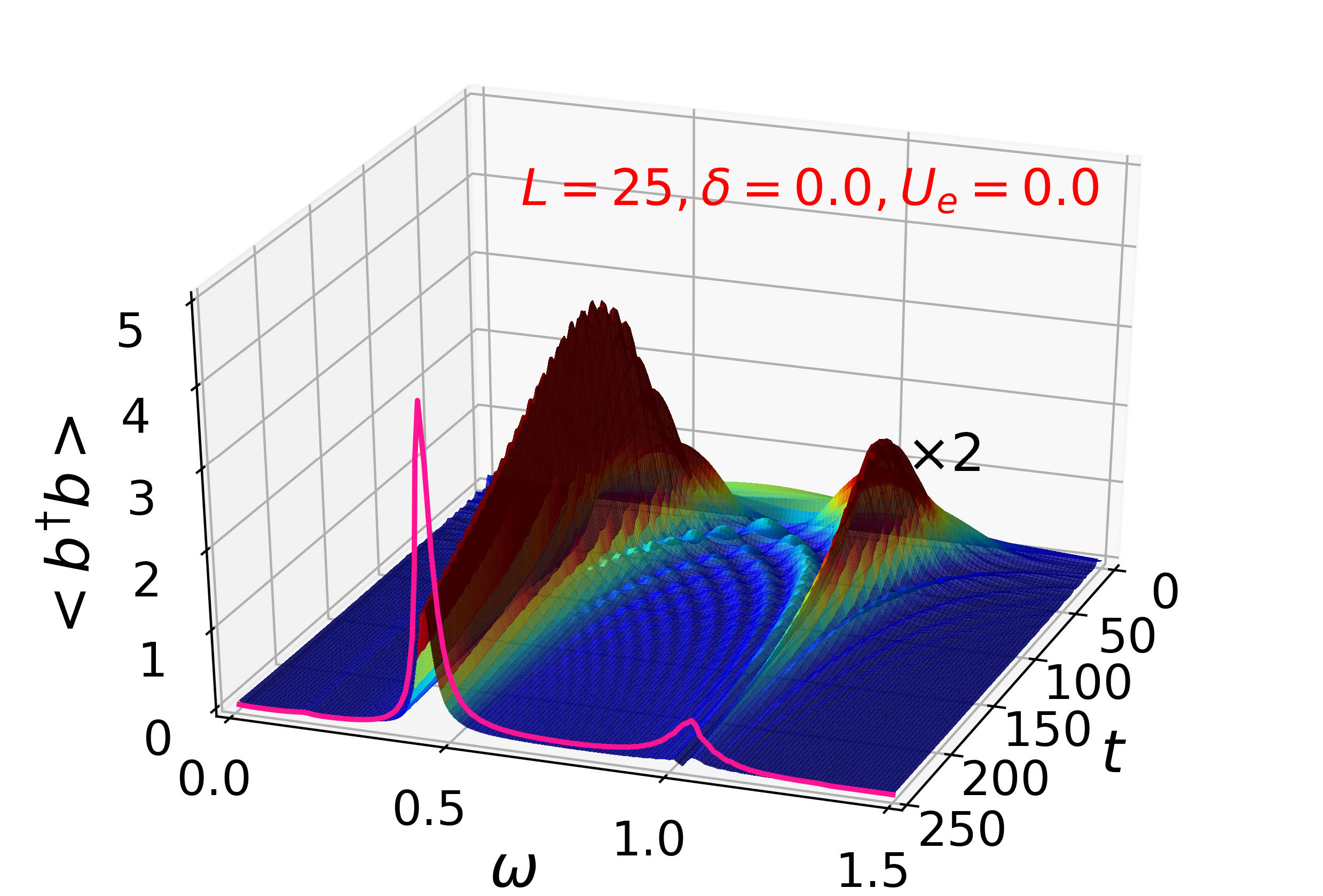}};
  \node at ( 6.0,-2.90) {\includegraphics[width=0.33\columnwidth]{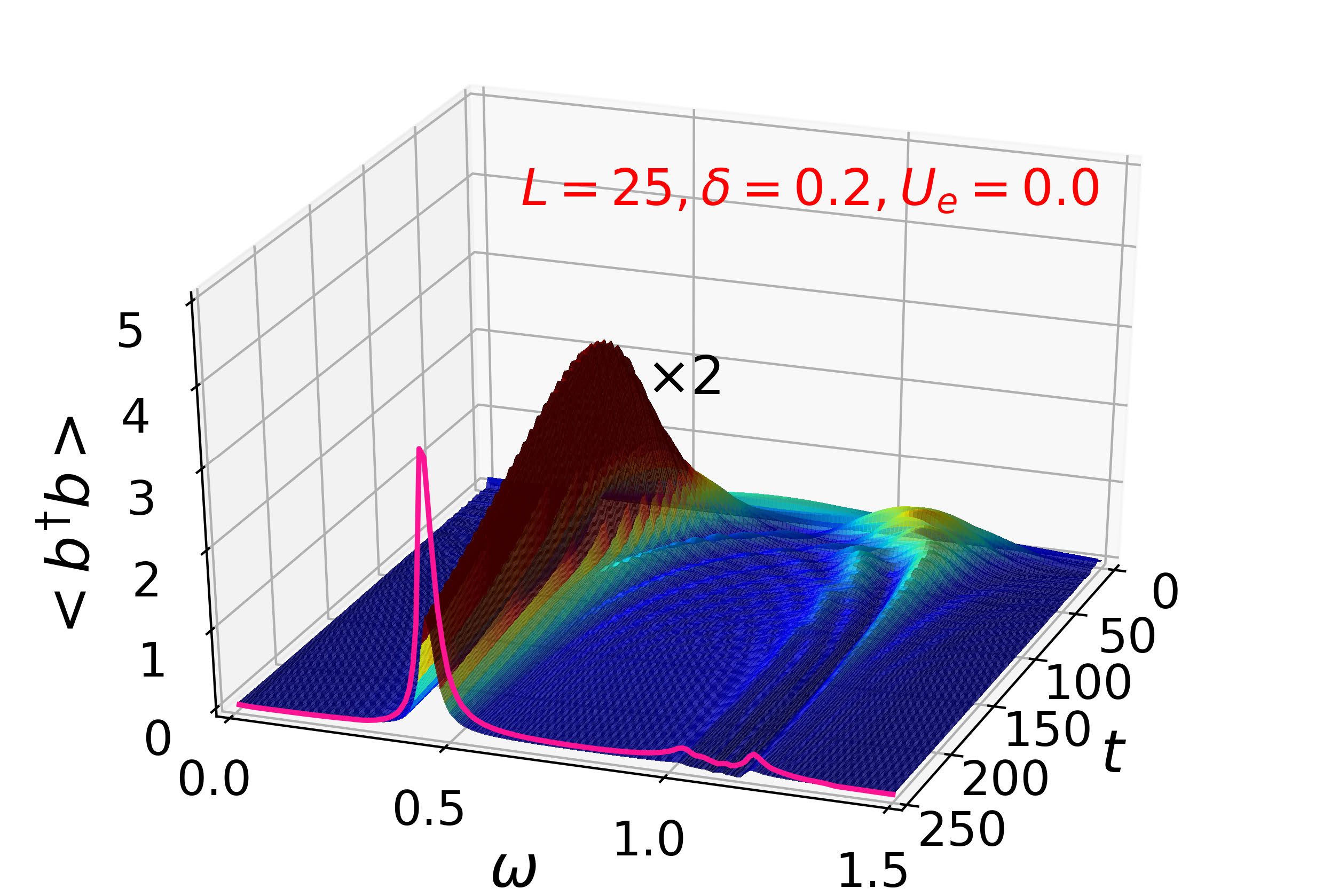}};
  \node at (12.0,-2.90) {\includegraphics[width=0.33\columnwidth]{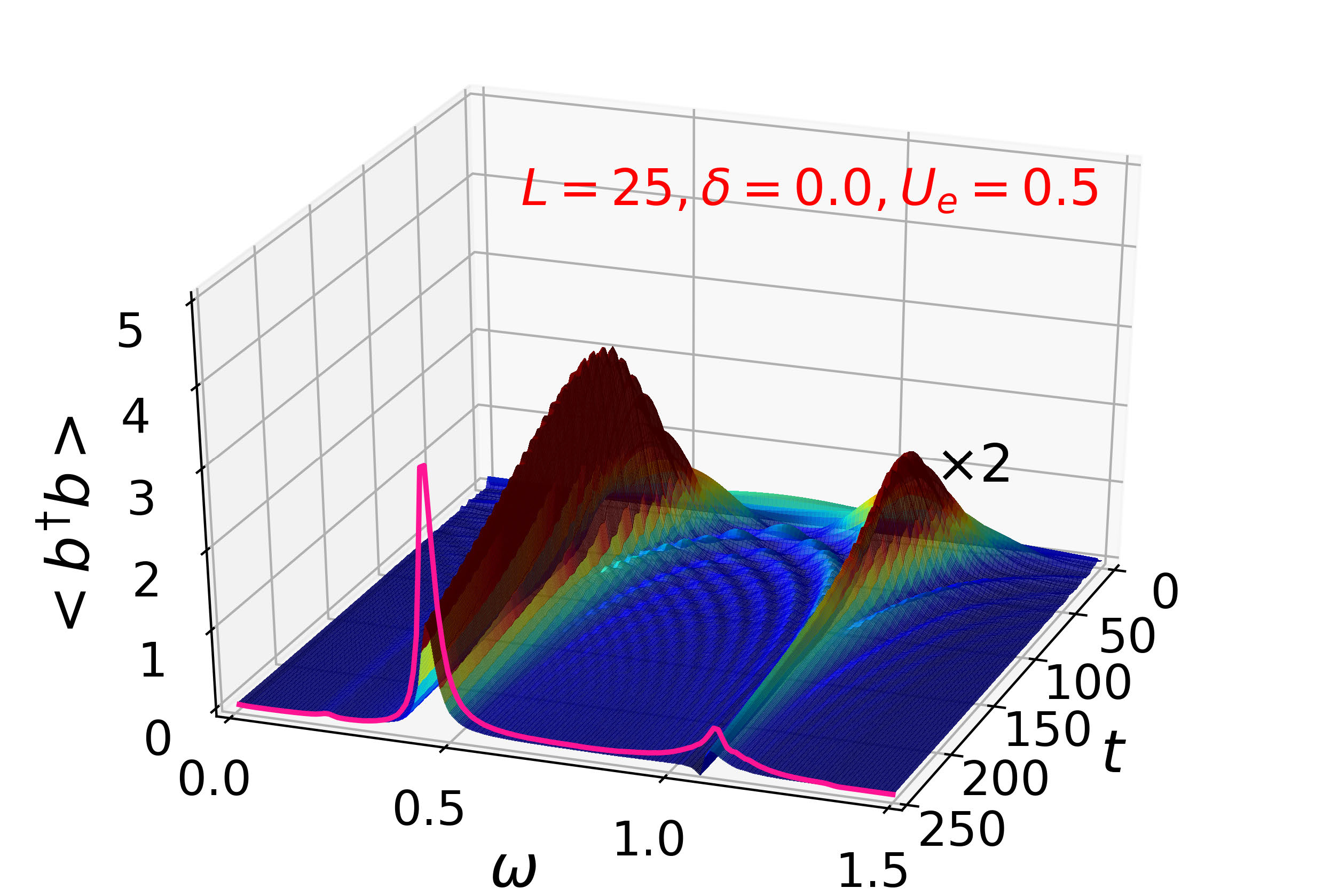}};
  \node[black] at (-1.5, 3.0) {$\bm{(a)}$};
  \node[black] at ( 4.5, 3.0) {$\bm{(b)}$};
  \node[black] at (10.6, 3.0) {$\bm{(c)}$};
  \node[black] at (-1.5,-1.80) {$\bm{(d)}$};
  \node[black] at ( 4.5,-1.80) {$\bm{(e)}$};
  \node[black] at (10.6,-1.80) {$\bm{(f)}$};
  \node[black] at (5.5, 4.0) {\Large{\bm{$L=15$}}};
  \node[black] at (5.5,-0.80) {\Large{\bm{$L=25$}}};
 \end{tikzpicture}
 \caption{The effect of cavity leakage. The time-dependent plots (the pink curves) refer to results with (without) the
   classical baths for $L=15$ (upper panels) and $L=25$ (lower panels).  The values of $\delta$ and $U_e$ used in a
   given panel are indicated at the top of the panel.  The bath parameters for the cavity leakage calculations are
   $A=0.005$, $a=0.6$, ${ \Delta_B} =0.01$ and $N_{bath}=200$. The time evolving plots are scaled by `$2$' for visual
   clarity.  }
 \label{fig6}
\end{figure*}
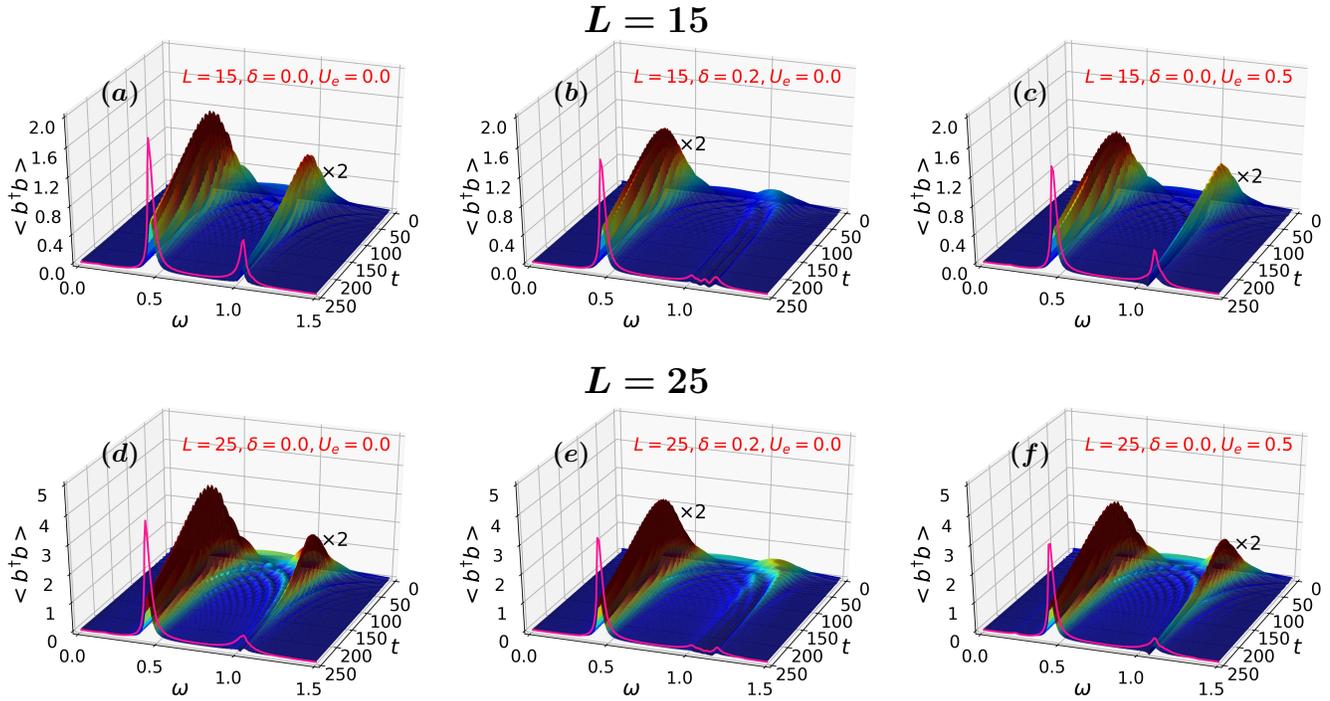

In the presence of cavity leakage, the NEGF-GKBA scheme in the Eqs.~(\ref{NEGF-GKBA}a-d) will remain the same except for
Eq.~\ref{NEGF-GKBA}(a){ \cite{Sakkinen2015}}, since now the time evolution of the average position will have a
contribution from the cavity leakage:
\begin{equation}
  i\frac{d}{dt}\phi_\mu=\sum_\nu h^b_{\mu\nu}\phi_\nu
  + \sum_{\nu,ij}\alpha_{\mu\nu}g_{\nu,ij}\rho^<_{ji}+\sum_\nu\alpha_{\mu\nu}F_\nu.
\end{equation}
Thus, together with the NEGF-GKBA time-evolution, the coordinates of the classical oscillators are time-evolved using
the Ehrenfest dynamics {\cite{Sakkinen2015,Bostrom2016}}. In the calculations, we considered all the oscillators' masses
$m_k=1$. Hence,
\begin{align}
{\ddot{q}_{a,k}}(t)&=\nobreak-\omega_k^2 {q_{a,k}}(t)+C_k\langle \hat a^\dagger+\hat a\rangle_{{\bar{q}_a,\bar{q}_b,t}},\\
{\ddot{{q}}_{b,k}}(t)&=\nobreak-\omega_k^2 {{q}_{b,k}}(t)+C_k\langle \hat b^\dagger+\hat b\rangle_{{ \bar{q}_a,\bar{q}_b,t}},
\end{align} 
where $\bar q_a = \{{q_{a,k}}\}$ and $\bar q_b = \{{q_{b,k}}\}$, also, we have considered that, at initial time, the
bath oscillators are in their equilibrium positions, i. e. $q_{a,k}=q_{b,k}=0$ and $p_{a,k}=p_{b,k}=0$.  The
effects of cavity leakage are shown in \textbf{Figure \ref{fig6}}, where the pink curve corresponds to the long-time
limit result without cavity leakage, and the time-evolving plots are obtained by including cavity leakage (the values of
the parameters used in the calculations are reported in the figure caption).  In all the cases shown, the time-evolved
plots are less intense than the corresponding pink curve. That is, cavity leakage reduces the intensity of the
fluorescent spectra also in the presence of disorder and interaction. This is expected, since leakage removes photons
from the optical cavity.
 
\section{Third harmonic generation}\label{terza}
\begin{figure}
	\begin{tikzpicture}
	\node at (0.0, 2.3) {\includegraphics[width=0.5\columnwidth]{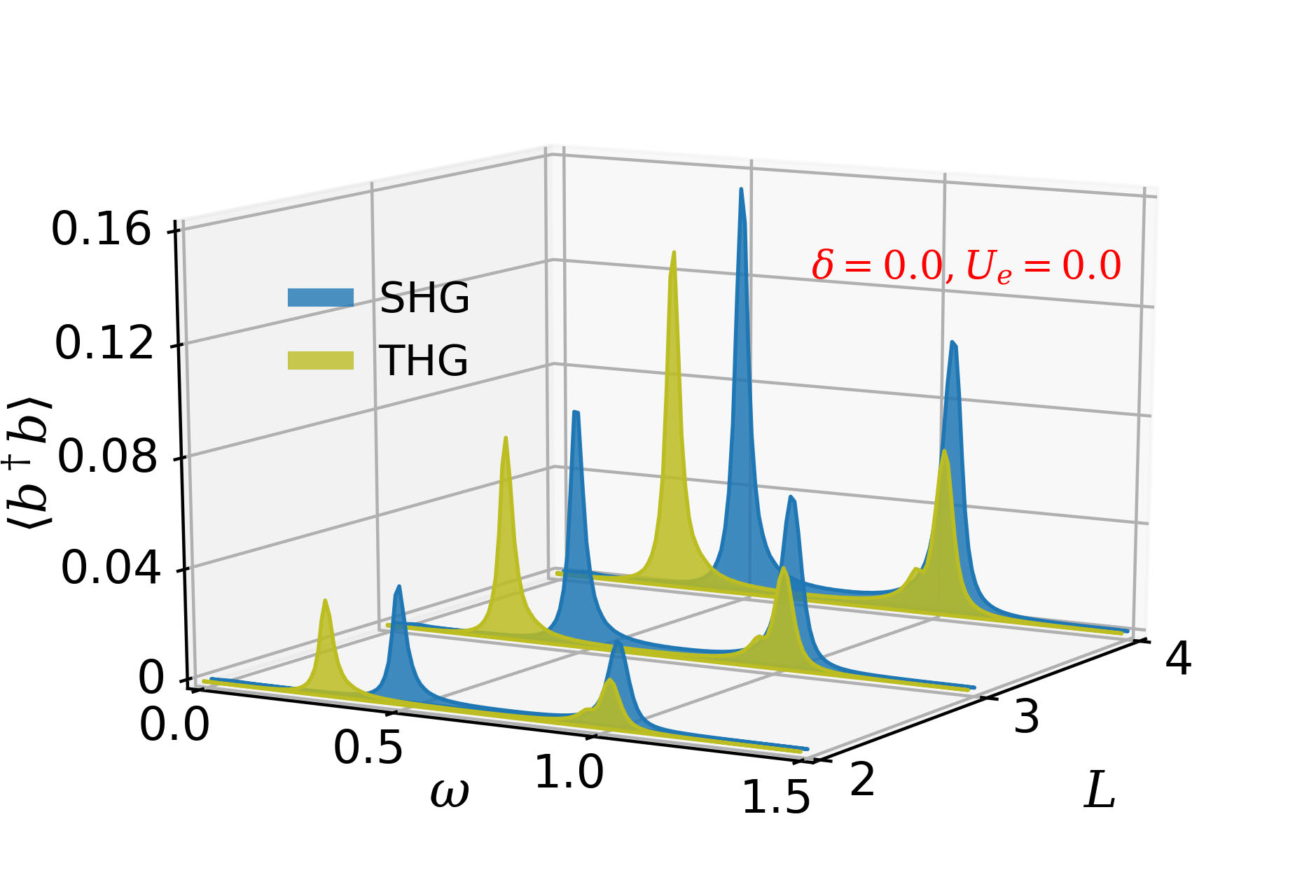}};
	\node at (9.0, 2.3) {\includegraphics[width=0.5\columnwidth]{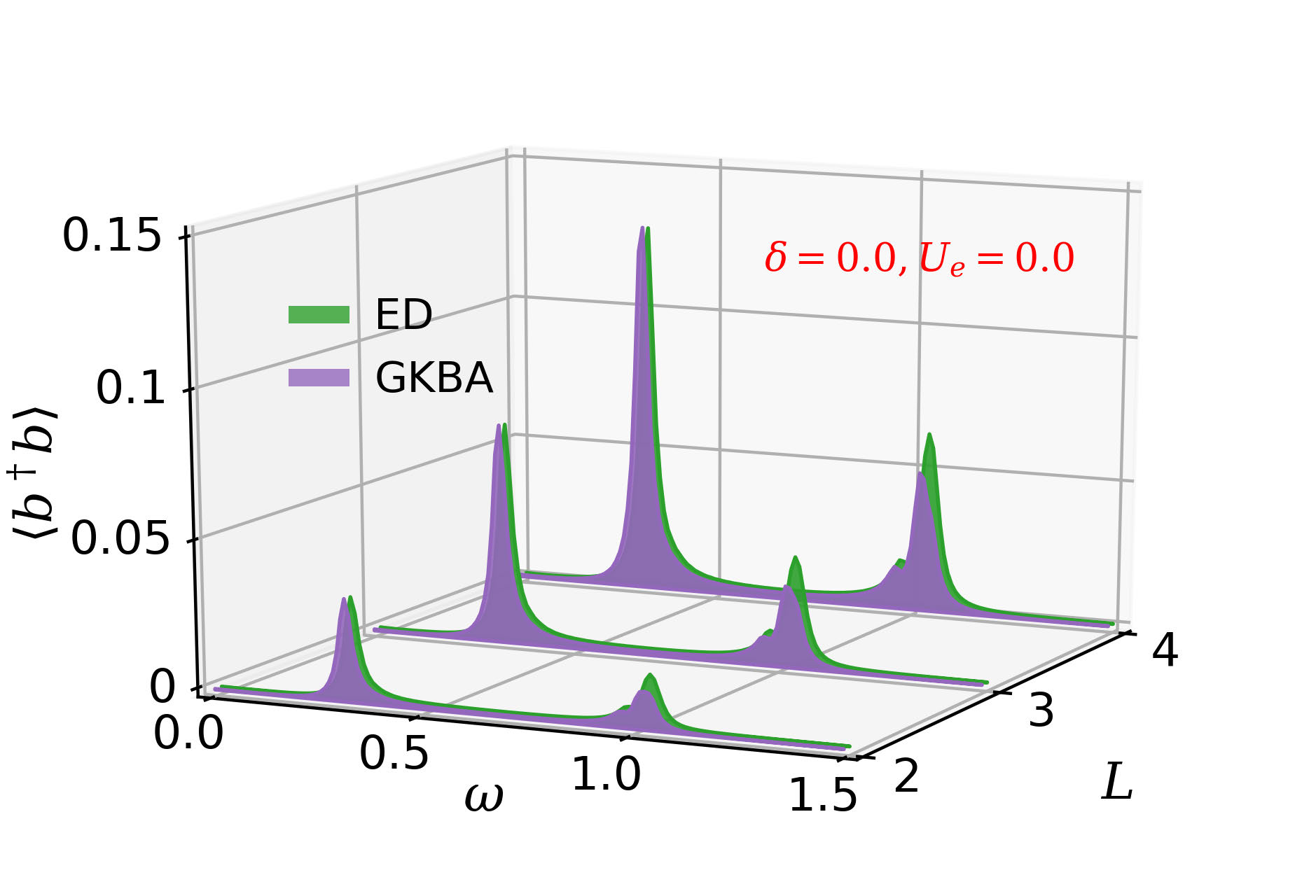}};
	\node[black] at (-0.40,4.) {\large$\bm{(a)}$};
    \node[black] at (8.40,4.) {\large$\bm{(b)}$};
	\end{tikzpicture}
	\caption{ Panel $\bm{(a)}$: comparison of fluorescent spectra for SHG and THG obtained with ED. Panel
          $\bm{(b)}$: comparison between THG fluorescent spectra obtained with ED and NEGF-GKBA. All the SHG (THG)
          calculations in figure are with ${\omega_a}={\Delta}/2=1/2$ (${\omega_a}={\Delta}/3=1/3$). For both panels,
          $U_e=\delta=0.0$, with the spectra obtained at time $t_e=250$.}
	\label{fig7}
\end{figure}

All the results so far were for the second harmonic spectrum, i.e. we considered ${\omega_a}={\Delta}/2$. Clearly, the
same NEGF-GKBA (and ED) treatments can be employed to study third harmonic generation (THG), with
${\omega_a}={\Delta}/3$. In comparison to SHG, the THG peaks are less intense, as it can be observed in \textbf{Figure
  \ref{fig7}} $\bm{(a)}$, where the results were obtained with ED. We then compared THG spectra obtained with the ED and
NEGF-GKBA in the case of non-interacting, non-disordered system. As shown in \textbf{Figure \ref{fig7}}$\bm{(b)}$, there
is a quite good agreement between the results from the two methods.

\begin{figure}
\centering
	\begin{tikzpicture}
	\node at (-2.3, 2.3) {\includegraphics[width=0.48\columnwidth]{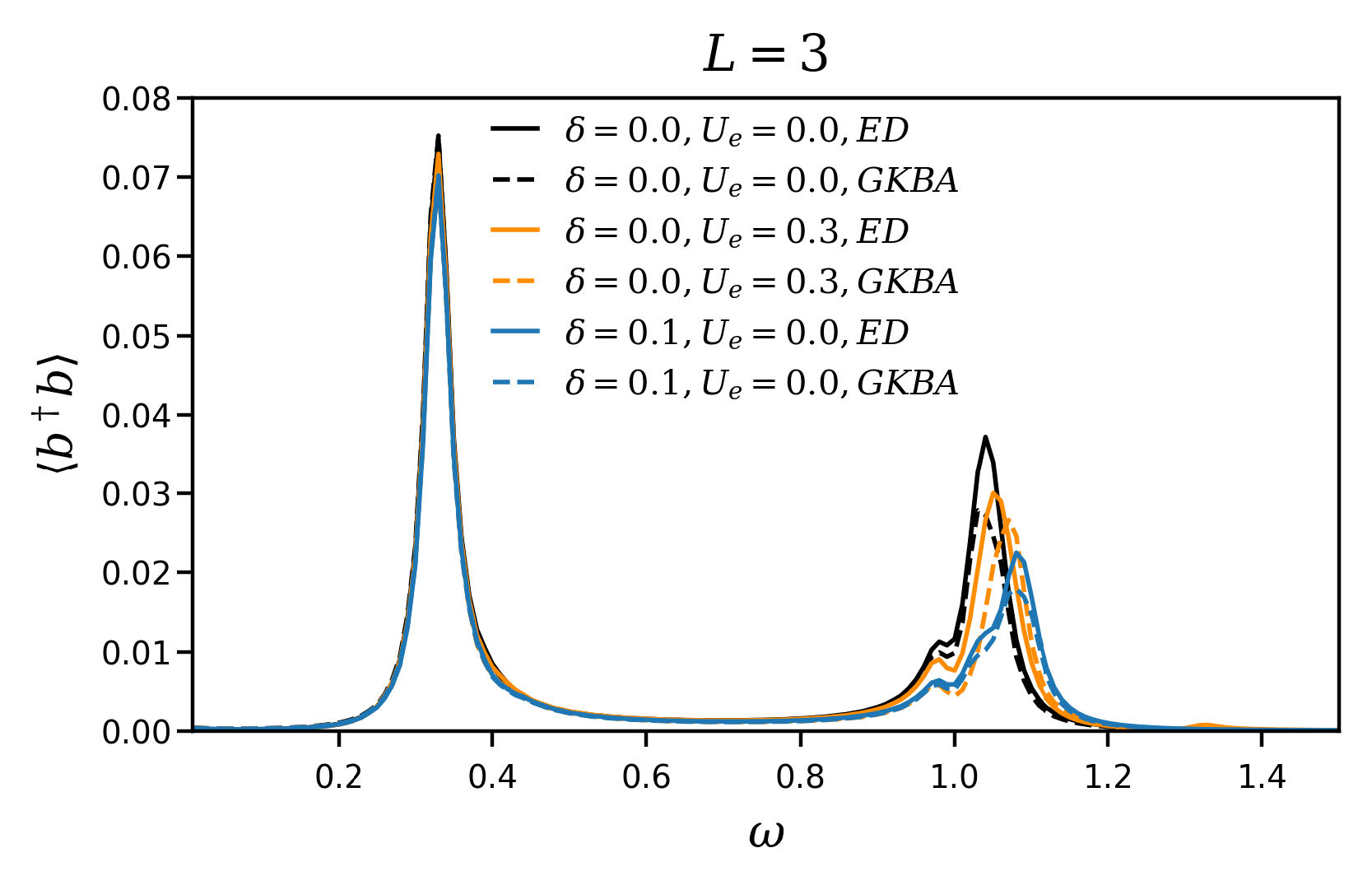}};
	\node at (7.3, 2.3){\includegraphics[width=0.48\columnwidth]{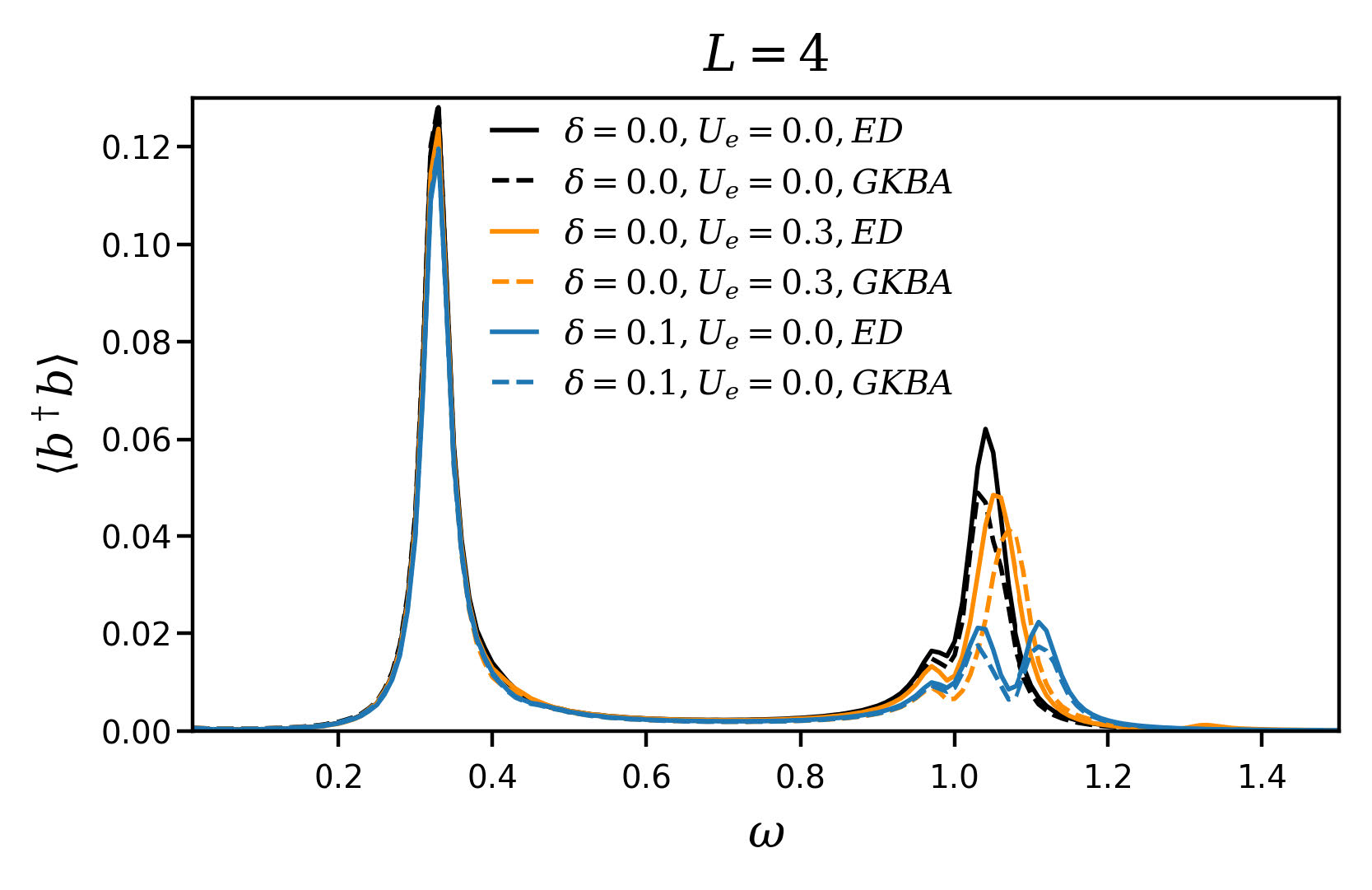}};
	    \node[black] at (1.20,4.0) {\large$\bm{(a)}$};
     	\node[black] at (10.50,4.0) {\large$\bm{(b)}$};
	\end{tikzpicture}
	\caption{ Comparison between THG fluorescent spectra from ED and NEGF-GKBA treatments,
         and with or without disorder and electron interactions (the used values of 
         $\delta$ and $U_e$ are shown in the legends).The results are for small systems ($L=3, 4$),
         and the spectra are obtained at time $t_e=250$.    } 
	\label{fig8}
\end{figure}

Finally, THG results from NEGF-GKBA (and their comparisons to ED results) when disorder and interaction are included are
shown in \textbf{Figure \ref{fig8}}. Again, the agreement between the two approaches is quite satisfactory. Also, as for
SHG, both disorder and interactions will reduce the intensity of THG (the damping of THG observed with the disorder
present is larger than with the interaction). As observed already for the case of SHG, the THG peak obtained with
NEGF-GKBA for $U_e=0.3$ is slightly shifted compared to the ED peak. We expect the source of this discrepancy to be same
as for SHG.  In these very preliminary discussion of THG, we examined only very small Dicke systems. However, from what
was found, and as suggested by the picture emerged for SHG, it would appear that the NEGF-GKBA method can be
successfully used to study THG spectra in large Dicke systems, and in the presence of disorder and interaction.

\section{Conclusions}\label{finito}
The Dicke model is one of the paradigmatic system in quantum optics to address nonlinear light-matter interaction
quantum phenomena. In this paper, we mainly studied the second-harmonic generation (SHG) fluorescent spectra of the
Dicke model in the presence of disorder and electron interaction. To describe large disordered and interacting samples,
we proposed the use of nonequilibrium Green's functions (NEGF) in the so-called generalized Kadanoff-Baym ansatz
(NEGF-GKBA) for electron-boson systems, a state-of-art approach based on nonequilibrium many-body perturbation theory.

The scope of the approach to describe SHG in the Dicke model was tested by comparing NEGF-GKBA and exact diagonalization
(ED) spectra for small interacting disordered systems and larger clean noninteracting ones. For the parameter regimes
investigated, we observed a very good agreement between the ED and NEGF-GKBA treatments (though, on increasing the
electron-electron interactions, the SHG peaks are slightly shifted in NEGF-GKBA as compared to ED). These findings point
at NEGF-GKBA as a powerful, rather accurate and viable method to deal with quantum systems of some
complexity. Furthermore

Concerning specific physical trends, our study suggests that both disorder and electron-electron interactions cause a
reduction in SHG intensity in the Dicke model, and that disorder hinders SHG more strongly than interaction. We also
considered the role of cavity leakage by coupling the photon fields to baths of classical oscillators. The results
showed (or, more properly speaking, confirmed) that, in accordance with one's intuition, cavity leakage depresses the
intensity of the fluorescent spectra.

Further, we briefly explored the application of the approach to the investigation of third harmonic generation (THG). We
considered only small samples, but these endorsed quite clearly the use of the method for THG. Again, for THG, results
from ED and GKBA agree very well, and similarly to SHG, the THG signal decreases with disorder and interactions. That
is, the method can be applied also to the study of higher harmonic generation processes.

Overall, our outlook expectation (and in part speculation) is that the NEGF-GKBA approach is generally robust to deal
with non linear phenomena in the Dicke model and/or related models of quantum optics (and their generalisations), { and
  that a mean-field treatment of e-e correlations is a useful starting point. However, to deal with degenerate states
  and range of parameters in the vicinity of energy level crossings, higher order diagrammatic contributions are to be
  included in the NEGF-GKBA self-energies.}

\medskip

\textbf{Funding information:} { This research performed in this work was part of projects funded by the Swedish Research
  Council with grants 2017-03945 and 2022-04486 (M.G. and C.V.)  This research was part of project
  no. 2021/43/P/ST3/03293 cofunded by the National Science Centre and the European Union’s Horizon 2020 research and
  innovation programme under the Marie Sklodowska-Curie grant agreement no. 945339 (Y.P.).}

\medskip
%

\begin{thebibliography}{10}
\providecommand{\url}[1]{\texttt{#1}}
\providecommand{\urlprefix}{URL }

\bibitem{BloeRMP}
N.~Bloembergen,
\newblock \emph{Rev. Mod. Phys.} \textbf{1982}, \emph{54}, 3 685.

\bibitem{physi}
M.~F. Ciappina, J.~A. P{\'e}rez-Hern{\'a}ndez, A.~S. Landsman, W.~A. Okell,
  S.~Zherebtsov, B.~F{\"o}rg, J.~Sch{\"o}tz, L.~Seiffert, T.~Fennel,
  T.~Shaaran, et~al.,
\newblock \emph{Rep. Prog. Phys.} \textbf{2017}, \emph{80}, 5 054401.

\bibitem{engine}
X.~Fu, T.~J. Cui,
\newblock \emph{Progress in Quantum Electronics} \textbf{2019}, \emph{67}
  100223.

\bibitem{chemist}
C.~Andraud, O.~Maury,
\newblock \emph{Eur. J. Inorg. Chem.} \textbf{2009}, \emph{2009}, 29-30 4357.

\bibitem{biolog}
S.~Yue, M.~N. Slipchenko, J.-X. Cheng,
\newblock \emph{Laser \& photonics reviews} \textbf{2011}, \emph{5}, 4 496.

\bibitem{medicine}
G.~F. Combes, A.-M. Vu{\v{c}}kovi{\'c}, M.~Peri{\'c}~Bakuli{\'c}, R.~Antoine,
  V.~Bona{\v{c}}i{\'c}-Koutecky, K.~Trajkovi{\'c},
\newblock \emph{Cancers} \textbf{2021}, \emph{13}, 16 4206.

\bibitem{Cini1993}
M.~Cini, A.~d'Andrea, C.~Verdozzi,
\newblock \emph{Phys. Lett. A} \textbf{1993}, \emph{180}, 6 430.

\bibitem{Cini1995}
M.~Cini, A.~d’Andrea, C.~Verdozzi,
\newblock \emph{Int. J. Mod. Phys. B} \textbf{1995}, \emph{9}, 10 1185.

\bibitem{Emil2020}
E.~Vi\~{n}as Bostr\"{o}m, A.~D'Andrea, M.~Cini, C.~Verdozzi,
\newblock \emph{Phys. Rev. A} \textbf{2020}, \emph{102}, 1 013719.

\bibitem{Dicke1954}
R.~H. Dicke,
\newblock \emph{Physical review} \textbf{1954}, \emph{93}, 1 99.

\bibitem{Kirton2019}
P.~Kirton, M.~M. Roses, J.~Keeling, E.~G. Dalla~Torre,
\newblock \emph{Advanced Quantum Technologies} \textbf{2019}, \emph{2}, 1-2
  1800043.

\bibitem{RWA1}
E.~Jaynes, F.~Cummings,
\newblock \emph{Proceedings of the IEEE} \textbf{1963}, \emph{51}, 1 89.

\bibitem{RWA2}
H.~J. Carmichael, D.~F. Walls,
\newblock \emph{J. Phys. B: At. Mol. Opt. Phys} \textbf{1976}, \emph{9}, 8
  1199.

\bibitem{TavisCummings}
M.~Tavis, F.~W. Cummings,
\newblock \emph{Phys. Rev.} \textbf{1968}, \emph{170}, 2 379.

\bibitem{Keeling2009}
J.~Keeling,
\newblock \emph{Phys. Rev. A} \textbf{2009}, \emph{79}, 5 053825.

\bibitem{Eastham2001}
P.~R. Eastham, P.~B. Littlewood,
\newblock \emph{Phys. Rev. B} \textbf{2001}, \emph{64}, 23 235101.

\bibitem{Tsyplyatyev2009}
O.~Tsyplyatyev, D.~Loss,
\newblock \emph{Phys. Rev. A} \textbf{2009}, \emph{80}, 2 023803.

\bibitem{ShuHeQPT2022}
S.~He, L.-W. Duan, Y.-Z. Wang, C.~Wang, Q.-H. Chen,
\newblock \emph{Phys. Rev. A} \textbf{2022}, \emph{106}, 3 033712.

\bibitem{PRLRydberg2013}
X.-F. Zhang, Q.~Sun, Y.-C. Wen, W.-M. Liu, S.~Eggert, A.-C. Ji, et~al.,
\newblock \emph{Phys. Rev. Lett.} \textbf{2013}, \emph{110}, 9 090402.

\bibitem{Kadanoff}
L.~Kadanoff, G.~Baym,
\newblock \emph{Quantum statistical mechanics {Green}'s function methods in
  equilibrium and nonequilibrium problems},
\newblock W.A. Benjamin, New York, \textbf{1962}.

\bibitem{Keldysh}
L.~V. Keldysh,
\newblock \emph{Sov. Phys. JETP} \textbf{1965}, \emph{20} 1018.

\bibitem{Balzer2012}
K.~Balzer, M.~Bonitz,
\newblock \emph{Nonequilibrium {Green}'s function approach to inhomogeneous
  systems},
\newblock Number 867 in Lecture notes in physics. Springer, Heidelberg,
  \textbf{2013}.

\bibitem{GSRvLbook}
G.~Stefanucci, R.~van Leeuwen,
\newblock \emph{Nonequilibrium {Many}-{Body} {Theory} of {Quantum} {Systems}:
  {A} {Modern} {Introduction}},
\newblock Cambridge University Press, Cambridge, \textbf{2013}.

\bibitem{Sakkinen2015}
N.~S\"{a}kkinen, Y.~Peng, H.~Appel, R.~van Leeuwen,
\newblock \emph{J. Chem. Phys.} \textbf{2015}, \emph{143}, 23 234102.

\bibitem{Lipavsky86}
P.~Lipavský, V.~\v{S}pi\v{c}ka, B.~Velický,
\newblock \emph{Phys. Rev. B} \textbf{1986}, \emph{34}, 10 6933.

\bibitem{Karlsson2021}
D.~Karlsson, R.~van Leeuwen, Y.~Pavlyukh, E.~Perfetto, G.~Stefanucci,
\newblock \emph{Phys. Rev. Lett.} \textbf{2021}, \emph{127}, 3 036402.

\bibitem{Pavlyukh_Mar_2022}
Y.~Pavlyukh, E.~Perfetto, D.~Karlsson, R.~van Leeuwen, G.~Stefanucci,
\newblock \emph{Phys. Rev. B} \textbf{2022}, \emph{105}, 12 125134.

\bibitem{Pavlyukh_Nov_2022}
Y.~Pavlyukh, E.~Perfetto, G.~Stefanucci,
\newblock \emph{Phys. Rev. B} \textbf{2022}, \emph{106}, 20 L201408.

\bibitem{Scipost}
M.~Gopalakrishna, E.~Vi{\~n}as~Bostr{\"o}m, C.~Verdozzi,
\newblock \emph{SciPost Physics} \textbf{2023}, \emph{15}, 4 138.

\bibitem{Schlunzen2020}
N.~Schl\"{u}nzen, J.-P. Joost, M.~Bonitz,
\newblock \emph{Phys. Rev. Lett.} \textbf{2020}, \emph{124}, 7 076601.

\bibitem{Joost2020}
J.-P. Joost, N.~Schl\"{u}nzen, M.~Bonitz,
\newblock \emph{Phys. Rev. B} \textbf{2020}, \emph{101}, 24 245101.

\bibitem{PhysRevB.105.125135}
Y.~Pavlyukh, E.~Perfetto, D.~Karlsson, R.~van Leeuwen, G.~Stefanucci,
\newblock \emph{Phys. Rev. B} \textbf{2022}, \emph{105}, 12 125135.

\bibitem{Caldeira1983}
A.~O. Caldeira, A.~J. Leggett,
\newblock \emph{Annals of physics} \textbf{1983}, \emph{149}, 2 374.

\bibitem{Venkataraman2014}
V.~Venkataraman, A.~D.~K. Plato, T.~Tufarelli, M.~S. Kim,
\newblock \emph{J. Phys. B: At. Mol. Opt. Phys} \textbf{2013}, \emph{47}, 1
  015501.

\bibitem{Bostrom2016}
E.~Bostr{\"o}m, M.~Hopjan, A.~Kartsev, C.~Verdozzi, C.-O. Almbladh,
\newblock \emph{J. Phys. Conf. Ser.} \textbf{2016}, \emph{696}, 1 012007.

\end{thebibliography}

\end{document}